\documentclass[pre, aps, preprint, superscriptaddress]{revtex4-1}
\pdfoutput=1
\usepackage{amsmath,natbib,amsfonts}
\usepackage{amssymb,xcolor}
\usepackage{graphicx}
\usepackage{subfig}
\usepackage{tikz}
\usepackage[colorlinks=true, citecolor=blue, urlcolor = blue, linkcolor= red,bookmarks=true]{hyperref}
\newcommand{\beqa}{\begin{eqnarray}}
\newcommand{\eeqa}{\end{eqnarray}} 

\newcommand{\ran}{\rangle}

\begin{document}
\title{Poisson-shot-noise induced escape dynamics of a passive particle and a dimer}
\author{Gemeit Goel}
\email{grsg82@gmail.com}
\affiliation{Department of Physics, Indian Institute of Technology Delhi, Hauz Khas 110016, New Delhi, India}
\author{Rahul Marathe}
\email{maratherahul@physics.iitd.ac.in}
\affiliation{Department of Physics, Indian Institute of Technology Delhi, Hauz Khas 110016, New Delhi, India}


\begin{abstract}
Mean First Passage Time (MFPT) is a crucial quantity that characterizes stochastic processes in various stochastic dynamical systems across different disciplines. The inverse of the MFPT, known as the escape rate, is helpful in the context of Brownian motion and chemical processes. The MFPT dynamics of a passive particle in the presence of active noises presents a fascinating avenue. In this work, we study the dynamics of two different systems: first, a single Brownian particle and second, a dimer (two Brownian particles connected by a spring). In both cases, the systems are kept in contact with a thermal bath and an active bath, and are trapped in a bistable potential. The active bath produces kicks that follow the Poisson distribution. The MFPT of the particle and the dimer is investigated in various scenarios. Analytical results are presented for the small kick variance approximation only. In this approximation, we formulate the MFPT in two scaling regimes: a weak-noise regime characterized by classical Arrhenius scaling for escape rate and a strong-noise regime in which the MFPT scales as a power law in the effective noise strength. For the general case, the problem is solved numerically. Next, under the same approximation, the MFPT of a dimer in the small-bond-strength regime is studied using a sequential-hopping mechanism, and in the large-bond-strength regime, the MFPT is formulated using the adiabatic elimination approach. Motivated by recent studies on speeding up the escape mechanism, the MFPT of the particle and the dimer in the presence of an intermediate barrier is also examined. We deduced from the numerical results that the MFPT can be enhanced and further optimized by selecting a suitable intermediate potential profile.

\end{abstract}


\maketitle



\section{Introduction}

The average time a system takes to cross a barrier, known as the Mean First Passage Time (MFPT), is of central importance in a wide range of physical, chemical, and biological systems \cite{metzler2014first}. For example, in protein folding, of particular interest is the timescale over which the polypeptide chain converges to one folded state across the energy landscape \cite{lee2003first, polizzi2016mean}, in chemical reactions, the reaction rate is an important parameter which quantify the mean time for reactants to overcome an activation barrier to produce products \cite{kramers1940brownian}, other examples include statistical fluctuation in lasers \cite{zhu1986statistical}, socio-economic systems \cite{masoliver2014first}, etc. One key quantity in the context of Brownian motion and chemical kinetics is the escape rate which can be defined as the inverse of the MFPT when the system equilibrates much faster than the MFPT. In 1940, H. A. Kramers was the first to derive a relation for the escape rate of Brownian particles caught in a potential well and subjected to irregular forces by the surrounding medium (thermal motion) to yield a model that elucidates the rates of chemical reactions \cite{kramers1940brownian}. In the limit of weak noise, Kramers showed that the Brownian particle's escape rate follows Arrhenius scaling and depends upon the potential barrier height and the curvature at the stable and unstable points. In the years that followed, Kramers' work has been extended to several interesting systems \cite{hanggi1990reaction, mel1986theory}. Extensive studies have been done on the escape rate of Brownian particles in delta-correlated white noise \cite{singh2017escape, porra1993bistability} and colored noise \cite{hanggi1984bistable, lyngdoh2024escape, masoliver1987bistability, ZacconePRE2017}. Kramers' theory provides an understanding of crossing events driven by equilibrium thermal fluctuations. However, many biological and synthetic systems \cite{zhang2017janus, han2025field} operate far from equilibrium, where internal fluctuations give rise to active dynamics. The question of interest arises to discuss the escape rate problem in the regime of active systems \cite{geiseler2016kramers, sharma2017escape, militaru2021escape, woillez2019activated, miao2025escape}. Escape rates have been studied using the non-Markovian yet Gaussian active noises \cite{ray2007generalization,schuller2020kramers, farah2021exponentially, bicout1998escape, lyngdoh2024escape},  however, the dynamics in the presence of real non-Gaussian active noises remains largely unexplored for example in presence of Poisson-shot noise (PSN) \cite{woillez2019activated, bao2005cancellation,baule2023exponential, gera2021solution, han2015first}. Historically, shot noise was first discussed by  W. Schottky to explain the spontaneous fluctuation in the electric current \cite{schottky1918spontane}. Shot noise has also been observed experimentally and analyzed theoretically in the gating current of the population of sodium channels \cite{conti1989quantal, crouzy1993fluctuations}. Unlike a thermal bath, the displacement distribution of a colloid in a bacterial suspension is non-Gaussian and heavy-tailed \cite{krishnamurthy2016micrometre, roy2021tuning, chen2024enhanced}. PSN is a natural candidate for modeling such non-equilibrium systems with non-Gaussian noise statistics.

\begin{figure}[!t]
    \centering
    \includegraphics[width=\textwidth]{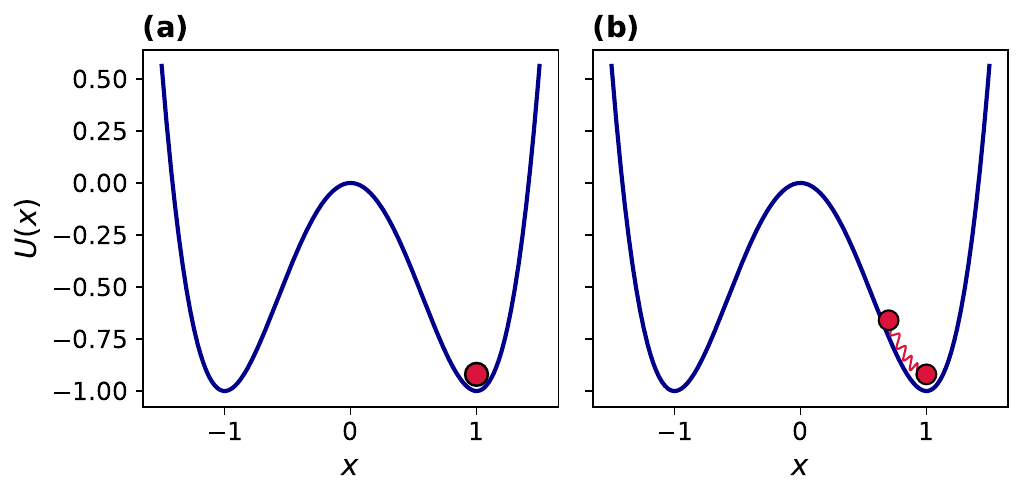}
    \caption{ \textbf{(a)} A Brownian particle trapped in a bistable potential well. \textbf{(b)} A dimer trapped in a bistable potential well. In both the cases particles experience thermal fluctuations (thermal bath) and  active kicks (active bath) modeled as the Poisson Shot Noise (PSN).}
    \label{fig:bistable_potential}
\end{figure}

A dilute solution of active particles, for example, bacteria, can impart kicks to the suspended passive particles. The statistics of these kicks can be modeled using PSN \cite{kingman1992poisson, park2020rapid, di2024brownian}. Some recent studies, motivated by the emerging field of active matter, examine the behavior of overdamped Brownian particles in contact with thermal and also PSN active baths \cite{di2024brownian}, and analyze escape from metastable states driven by non-Gaussian noise in the weak-noise limit \cite{baule2023exponential, gera2021solution}. The MFPT and the escape mechanism of the dimer are also of interest (see Fig. \ref{fig:bistable_potential}). Recent studies have examined the escape rate of the dimer under white and colored noise and find that the escape rate is highly dependent on the bond strength between the monomers and the noise parameters \cite{lyngdoh2024escape, asfaw2012exploring, ZacconePRE2017, singh2017kramers, lyngdoh2026escape}. In this paper, we study a single particle in contact with thermal and active baths and show that the Poisson-shot noise can be approximated by thermal noise in the small variance limit and discuss the general case through numerical results. In the small variance approximation, we also explore weak and strong noise regimes. The particle model is also generalized to study the dynamics of a dimer consisting of two harmonically coupled monomers interacting via a Morse potential. The sequential hopping mechanism and the adiabatic elimination method are employed to discuss the MFPT for weak and stiff bonds, respectively, under the small-variance approximation. 

Due to the exponential dependence of the escape rate on the barrier height in the weak noise limit for Gaussian processes, higher barriers have extremely low escape rates. Therefore, optimization of the escape rate is an important aspect to study. This is achieved in several ways, for example, by resonant activation: where the potential profile fluctuates and the escape rate exhibits a local maxima at the resonant fluctuating frequency \cite{doering1992resonant, mantegna2000experimental}; using noise engineering, barrier crossing can be accelerated for the colored noise by selecting a suitable non-Markovian memory kernel \cite{kappler2018memory}; escape rate can also be enhanced using resetting protocols \cite{singh2020resetting, pal2024random, pal2024channel, ahmad2019first, capala2023optimization, ray2020diffusion}; escape rate can enhance dramatically by potential roughness showing peak at optimal roughness amplitude \cite{fan2026hindrance}. Introduction of an additional barrier can also significantly enhance the particle's escape rate \cite{kumar2025inferring, kumar2025speeding, chupeau2020optimizing}. Motivated by these studies, we also investigate the effect of an intermediate barrier on the MFPT and the escape rate of a single particle and a dimer in the presence of passive and active noises. The simulation results show that the escape rate increases for a Brownian particle and a dimer in the presence of an intermediate potential. Interestingly, the intermediate potential can be tuned by varying the curvature and the intermediate barrier height to further optimize the MFPT.

The rest of the paper is organized as follows. In section \ref{sec: A single Brownian particle}, the escape of a single Brownian particle is studied, its statistical features are discussed, and the MFPT and the escape rate of the particle in the weak noise regime and also the strong noise regime in the small kick-variance limit is studied. In section \ref{sec:dimer}, a model of a dimer is studied. It's escape rate is analytically and numerically analyzed in the small bond strength limit and the large bond strength limit. In section \ref{sec:intermediate-potential}, we discuss the effects of introducing an intermediate barrier on the escape rate through simulations and conclude in Sec. \ref{conclude}.

\section{A single Brownian particle}
\label{sec: A single Brownian particle}
The model consists of a one-dimensional overdamped Brownian particle trapped in a bistable potential. The particle is simultaneously in contact with a thermal bath and an active bath. If $x(t)$ is the instantaneous position of the particle. The dynamics of the particle is given by the overdamped Langevin equation,
\begin{equation}
\gamma\frac{dx(t)}{dt} = -U'(x)+\sqrt{2k_BT\gamma}~\xi(t) + \gamma\eta(t),
\label{eq: overdamped langevin equation}
\end{equation}
where $\gamma$ is the friction coefficient, $k_B$ is the Boltzmann constant, and $T$ is the absolute temperature of the thermal bath. The $\xi(t)$ is the thermal noise such that $\langle \xi(t) \rangle = 0$, and $\langle \xi(t)\,\xi(t') \rangle = \delta(t - t')$. The $\eta(t)$ is the active PSN that models the interaction of passive particles with active particles suspended in the bath \cite{di2024brownian}. The system parameter $\gamma$ controls the physical effects of the bath on the intensity of active kicks, and the product $\gamma\eta(t)$ results in a stochastic force experienced by the Brownian particle. Moreover, the two noises $\xi(t)$ and $\eta(t)$ are assumed to be independent. The potential $U(x)$, as shown in Fig. \ref{fig:bistable_potential}, is a bistable potential of the form
\begin{equation}
U(x) = -\frac{\Omega_B}{2}x^2 + \frac{\Omega_B}{4x^2_m}x^4.
\label{eq:bistable_potential_general_form}
\end{equation}
The $\Omega_B$ is the magnitude of the curvature of the potential at $x=0$. The stationary points of the potential are located at $x = 0$ and $x = \pm x_m $, and the maximum height of the potential barrier at $x = 0$ with respect to the minimum is $U_B = \frac{x^2_m \Omega_B}{4}$.
\begin{figure*}[!t]
    \centering
    \includegraphics[width=\textwidth]{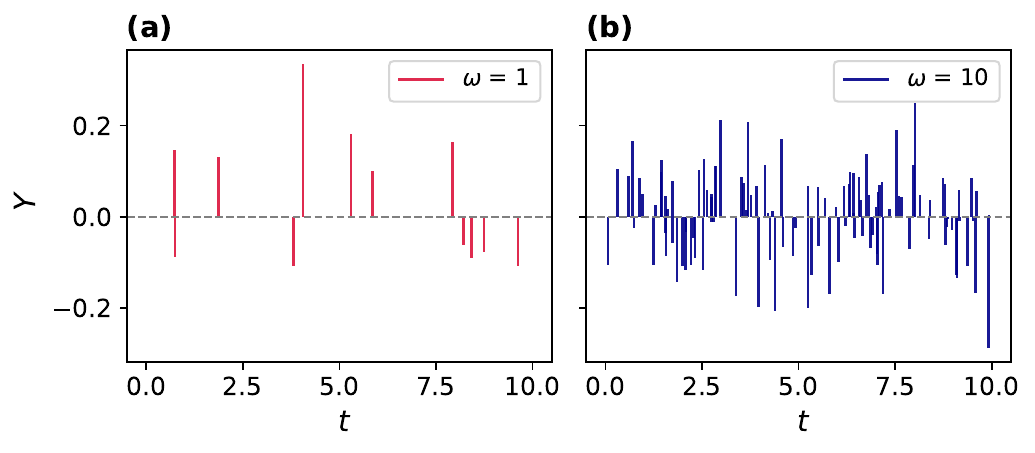}
    \caption{Plot of amplitude $Y$ of the kicks sampled from a Gaussian distribution Eq. (\ref{eq:gaussian_psn_kick}) with kick rate \textbf{(a)} $\omega = 1$ and \textbf{(b)} $\omega = 10$. We see that the increase in $\omega$ leads to an increase in the frequency of the kicks. Simulation time step is $\Delta t = 10^{-4}$.
}
    \label{fig:psn_kicks}
\end{figure*}
PSN can be written in the following form \cite{hanggi1980langevin}

\begin{equation}
\eta(t) = \sum_{j=0}^{N(t)}Y_j\delta(t-t_j).
\label{eq:PSN}
\end{equation}
Here, $N(t)$ is the number of kicks up to time $t$, and it is a Poisson random variable. As a consequence of Poisson-distributed events, the time between two successive kicks is exponentially distributed with rate $\omega$. Here, $Y_j$ represents the amplitude of the random kicks, and a sequence of them is independent and identically distributed. Kick amplitude $Y_j$ is sampled from a pre-defined probability distribution $\rho_a(y)$. The Probability distribution function of $N(t)$ is given by
\begin{equation}
P(n,t) = e^{-\omega t} \frac{(\omega t)^n}{n!}.
\label{eq:psn_poisson_distribtion}
\end{equation}
Poisson shot noise follows $\langle \eta(t) \rangle = \omega \langle Y \rangle$, and $\langle \eta(t) \eta(t^{\prime}) \rangle -  \langle \eta(t) \rangle \langle\eta(t^{\prime})\rangle= \omega \langle Y^2 \rangle \delta(t-t^{\prime}).$
Following \cite{di2024brownian}, the kick amplitude distribution $\rho_a(y)$ is chosen to follow a Gaussian distribution
\begin{equation}
\rho_a(y) = \frac{1}{Z} \exp\left(\frac{-\beta_a\gamma^2}{2m_a}y^2\right), 
\label{eq:gaussian_psn_kick}
\end{equation}
where $m_a$ is the mass of the active particle, and $\beta_a = \frac{1}{k_bT_a}$, $k_BT_a$ is the energy associated with the active particle, and the variance of the kicks is $\sigma_a^2 = m_a/\beta_a\gamma^2$, $\gamma$ is present in the distribution for dimensional consistency only. Here, $Z$ is the normalization constant. Fig. \ref{fig:psn_kicks} shows two realizations of the PSN with kick amplitudes sampled from the Gaussian distribution Eq. (\ref{eq:gaussian_psn_kick}), for two different kick rates $\omega$. This analysis can be generalized to accommodate different distributions for $\rho_a(y)$ (see Appendix \ref{nongauss}).


\subsection{Dimensionless form of the Langevin equation}
\label{sec:particle}
We now write the Langevin Eq. (\ref{eq: overdamped langevin equation}) in dimensionless form to make the model more general and to facilitate experimental validations. The barrier height, $U_B = \frac{x^2_m \Omega_B}{4}$, of the bistable potential shown in Fig. \ref{fig:bistable_potential} is used as the reference for the energy scale. Using $t_c$ and $x_m$ as reference scales for time and distance, the dimensionless variables are given by: $\bar{t} = \frac{t}{t_c}$, where, $t_c = \frac{\gamma x^2_m}{U_B}$, ~$\bar{x} = \frac{x}{x_m}$, $\bar{D} = \frac{k_BT}{U_B}$, ~$\overline{\xi(\bar{t})} = \xi(t) \sqrt{t_c}$, ~$\bar{\beta_a} = \beta_a U_B$,  $\bar{m_a} = \frac{m_a U_B}{\gamma^2x^2_m}$, $\bar{\omega} = \omega t_c$ and $\overline{\eta(\bar{t})} =  \eta(t) \gamma x_m/U_B$.  The typical experimental values of these parameters are $U_B  = 4.81 k_BT$, let $T = 25^\circ \mathrm{C}$, $x_m = 0.25 \mu \mathrm{m}$ and $ \gamma \approx 9.9 \times10^{-8} N ~sec/m$, adopted from the experimental study of transition path time in viscous medium of silica beads in bistable optical potentials \cite{ferrer2024transition}. For these reference parameters, the dimensionless parameters can be written as $\bar{t} = t/0.31$, $\bar{x} =x/0.25$ where unit of $x$ is $\mu m$, $\bar{D} = 0.21$, $\bar{\omega} = 0.31\omega$ and standard deviation of Gaussian kicks is $\bar{\sigma_a} = \sigma_a/0.25$. Here $\sigma_a = \sqrt{\frac{m_a}{\beta_a \gamma^2}}$. A typical experimental value of the standard deviation is $\sigma_a = 0.28 \mu m$, which gives $\bar{\sigma_a} = 1.12$. A typical kick rate is $\omega = 135 ~s^{-1}$, giving $\bar{\omega} = 41.85$. The experimental values of standard deviation and kick rate are in accordance with the experimental work of the reference \cite{roy2021tuning} on a polystyrene particle trapped using optical tweezers in an artificially engineered non-Gaussian reservoir for the study of mesoscale heat engines. Using these, the Langevin Eq. (\ref{eq: overdamped langevin equation}), after dropping the over-bars for convenience, takes the following dimensionless form:
\begin{equation}
\frac{dx(t)}{dt} = 4x(t) - 4x^3(t)+\sqrt{2D}~\xi(t) + \eta(t).
\label{eq: dimensionless overdamped langevin PSN}
\end{equation}
The distribution of kick amplitudes, Eq.(\ref{eq:gaussian_psn_kick}), takes the following dimensionless form  with variance $\sigma^2_a = m_a/\beta_a$  (over-bars have been dropped for convenience) 
\begin{equation}
\rho_a(y) = \frac{1}{Z} ~\exp\left(\frac{-\beta_a}{2m_a}y^2\right) .
\label{eq:gaussian_psn_normalized_kick}
\end{equation}
For the simulation of Eq. (\ref{eq: dimensionless overdamped langevin PSN}), the parameter values $D = 0.2$, $m_a = 0.01$, and $\beta_a = 1$ are used unless otherwise stated. The details of the simulation and numerical schemes used are given in Appendix \ref{simulate}. 

\begin{figure*}[!htbp]
    \centering
    \includegraphics[width=0.9\textwidth]{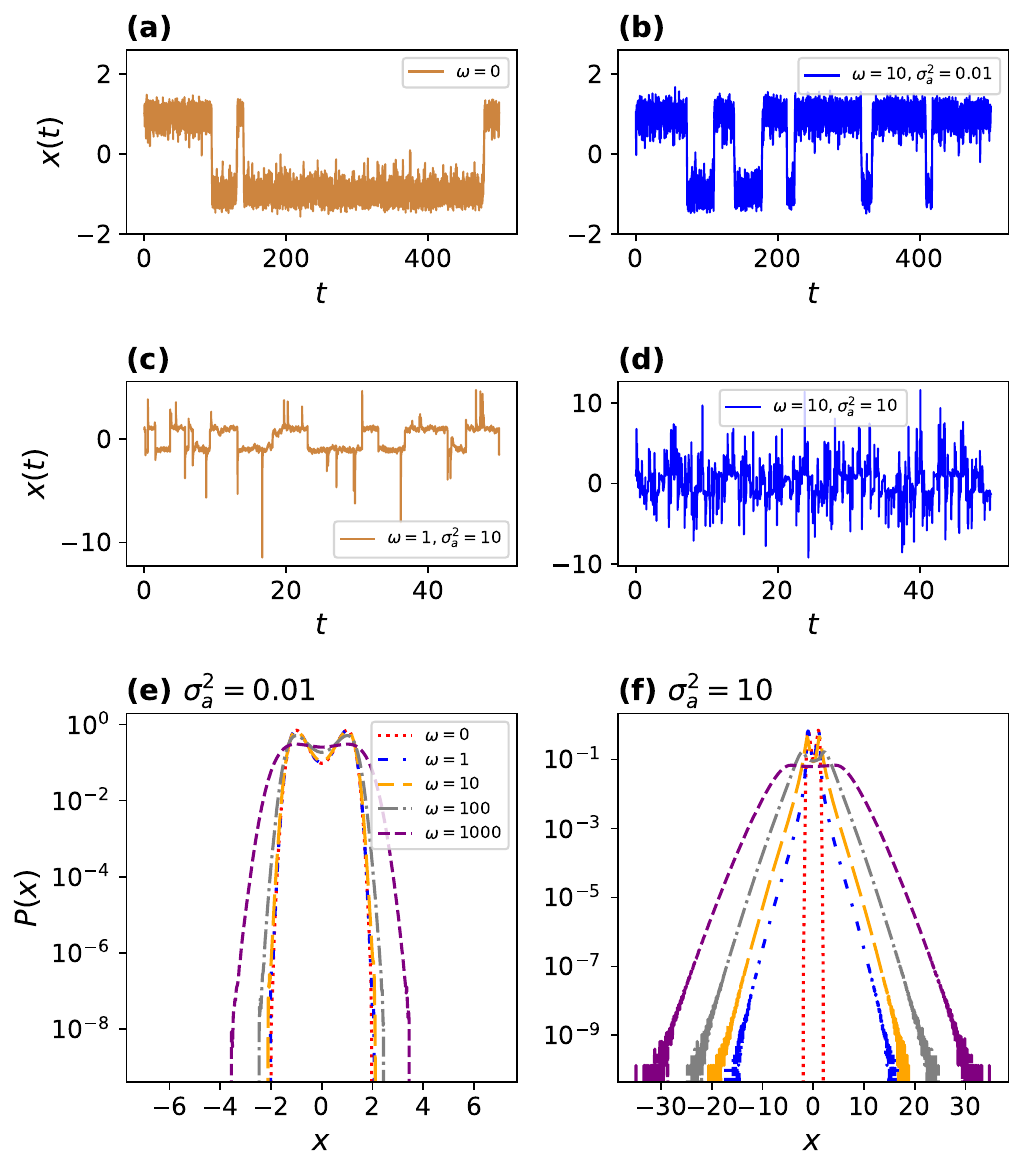}
    \caption{Trajectory of a Brownian particle in contact with a thermal bath and an active bath is shown: \textbf{(a)} kick rate $\omega = 0$ or only thermal bath, \textbf{(b)} $\omega =10,$ and variance $\sigma_a^2 =0.01$, \textbf{(c)} $\omega =1, \sigma_a^2 =10$, \textbf{(d)} $\omega =10, \sigma_a^2 =10$. Probability distribution plot of the position of a Brownian particle with $D = 0.5$, subjected to thermal noise and PSN with different variance and kick rates: \textbf{(e)} $\sigma_a^2 = 0.01$, \textbf{(f)} $\sigma_a^2 = 10$. As $\omega$ increases, hopping events increase, and therefore the head of $P(x)$ becomes flatten. Due to the large amplitude of Poissonian kicks, long tails in panel (f) can be observed. The simulation time step is $\Delta t = 10^{-5}$.}
    \label{fig:particle-trajectory-pdf}
\end{figure*}
The particle resides in one of the potential wells of the bistable potential and undergoes random motion within the well due to thermal and active fluctuations, and it hops to another well by crossing the potential barrier. The sample trajectories of such events for different parameters are shown in the first four panels of Fig. \ref{fig:particle-trajectory-pdf}.
\begin{figure*}[!t]
\centering
\includegraphics[width=\textwidth]{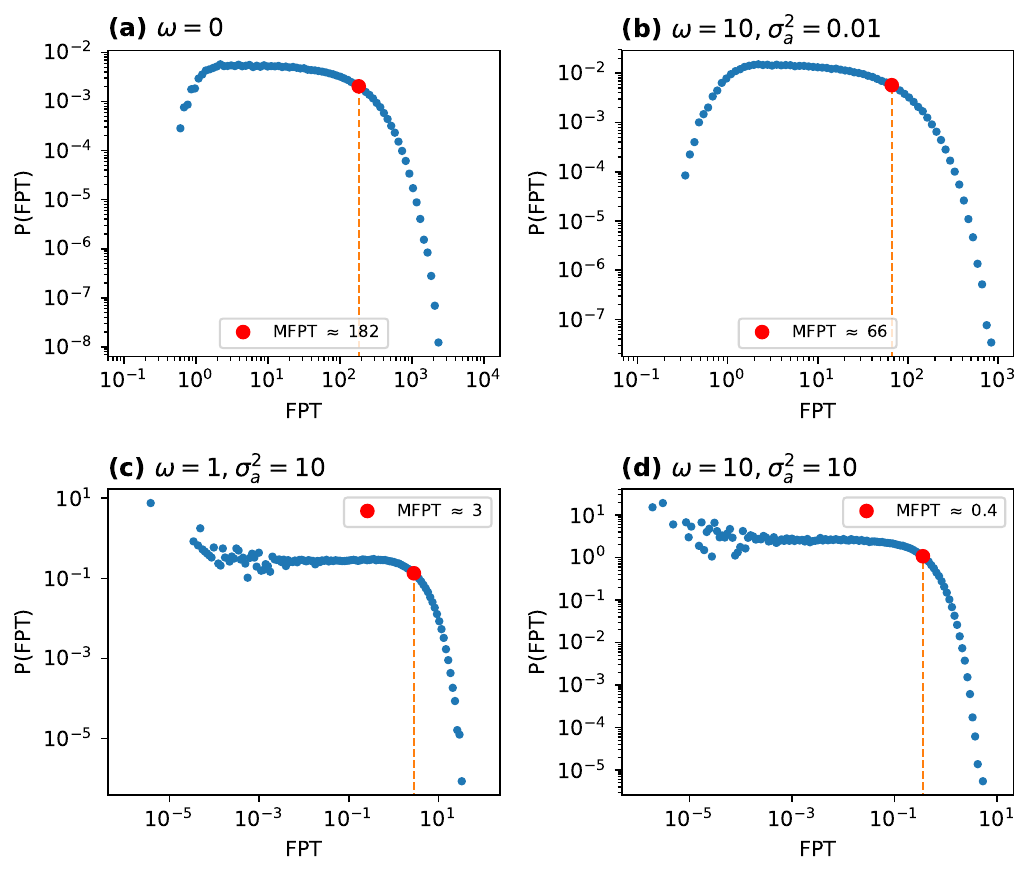}
\caption{Probability density for the First Passage Time (FPT) is plotted for different kick rates and variances. \textbf{(a)} $\omega = 0$, \textbf{(b)} $\omega = 10, \sigma_a^2 = 0.01$, \textbf{(c)} $\omega = 1, \sigma_a^2 = 10$, and \textbf{(d)} $\omega = 10, \sigma_a^2 = 10$. It can be observed that long tails are present in all cases and they shrink as we increase the variance of the kick amplitude, and for thermal noise and the small-variance case, events with short time scales relative to the mean time are much less probable. In contrast, for large variance, a saturation can be seen for small time scales. The time step for the simulation is $\Delta t = 10^{-4}$ for (a) and (b), and $10^{-6}$ for (c) and (d). A total of $5\times 10^4$ realizations were used for the plot.}
\label{fig:particle-FPT}
\end{figure*}
The probability distribution function (PDF) for the position of the particle is also shown in Fig. \ref{fig:particle-trajectory-pdf} panels (e) and (f). As expected, the PDF broadens upon the introduction of PSN, but long tails are significantly observed only for large $\sigma_a^2$. This observation gives the motivation for the small variance approximation. In the presence of only thermal noise, and when the barrier height is larger than the thermal energy,  $\Delta U > k_BT$, the Kramers' escape rate of a Brownian particle depends on the barrier height and the curvature at the maximum and minimum of the potential, given by
\begin{equation}
R = \frac{1}{2 \pi \gamma} \sqrt{|U''(x_{min})|U''(x_{max})|}~e^{\frac{-\Delta U}{k_BT}},
\label{eq: Kramer Escape Rate formula}
\end{equation}
where $U''(x_{min})$ and $U''(x_{max})$ are the curvatures of the potential at the minimum and maximum, respectively, and $\Delta U$ denotes the barrier height. By introducing dimensionless terms $\bar{R} = Rt_c$, $\bar{U''} = U''x_m^2/U_B$, and $\bar{\Delta U} = \Delta U/U_B$, the dimensionless form for the escape rate, after dropping the over-bars, can be written as
\begin{equation}
R = \frac{\sqrt{|U''(x_{min})|U''(x_{max})|}}{2\pi} e^{-\Delta U/D}.
\label{eq: Dimensionless form of Kramer formula}
\end{equation}
For the present case, the barrier height of the bistable potential is unity $\Delta U = 1,$ and the curvature at the maximum is $\Omega_B = -4$, and at the minimum it is $8$. Therefore, the escape rate can be written as
\begin{equation}
R = \frac{2\sqrt{2}}{\pi} e^{-1/D}.
\label{eq: Dimensionless form of Kramer escape rate for our case}
\end{equation}
 In case of Kramers' problem, we can define the escape rate $(R)$ as the inverse of the MFPT ($\mathcal{T}$), only if it is constant in the basin of the potential well, that is, the particle's equilibration time, say $\tau_{eq}$, is very small compared to the MFPT. But in the presence of PSN, the situation is non-equilibrium, and the system reaches a non-equilibrium steady state. From the work of reference \cite{di2024brownian}, the stationary state time scale for the particle to relax inside a harmonic potential well is $\tau_{ss} = \gamma/\Omega(x_m)$, where $\Omega(x_m)$ is the curvature at the minimum of the well. Therefore, for our case, we compare the $\mathcal{T}$ with $\tau_{ss}$, time scale for the particle to relax inside the potential well, $\tau_{ss} \approx \gamma/\Omega(x_m)$, where $\Omega_{x_m} = 2\Omega_B$.  In dimensionless form, it becomes $\tau_{ss} = 1/8$. Therefore, if $\mathcal{T} >> \tau_{ss}$, we can write the escape rate as the inverse of the MFPT.

\begin{figure}[!htbp]
\centering
\includegraphics[width=0.9\textwidth]{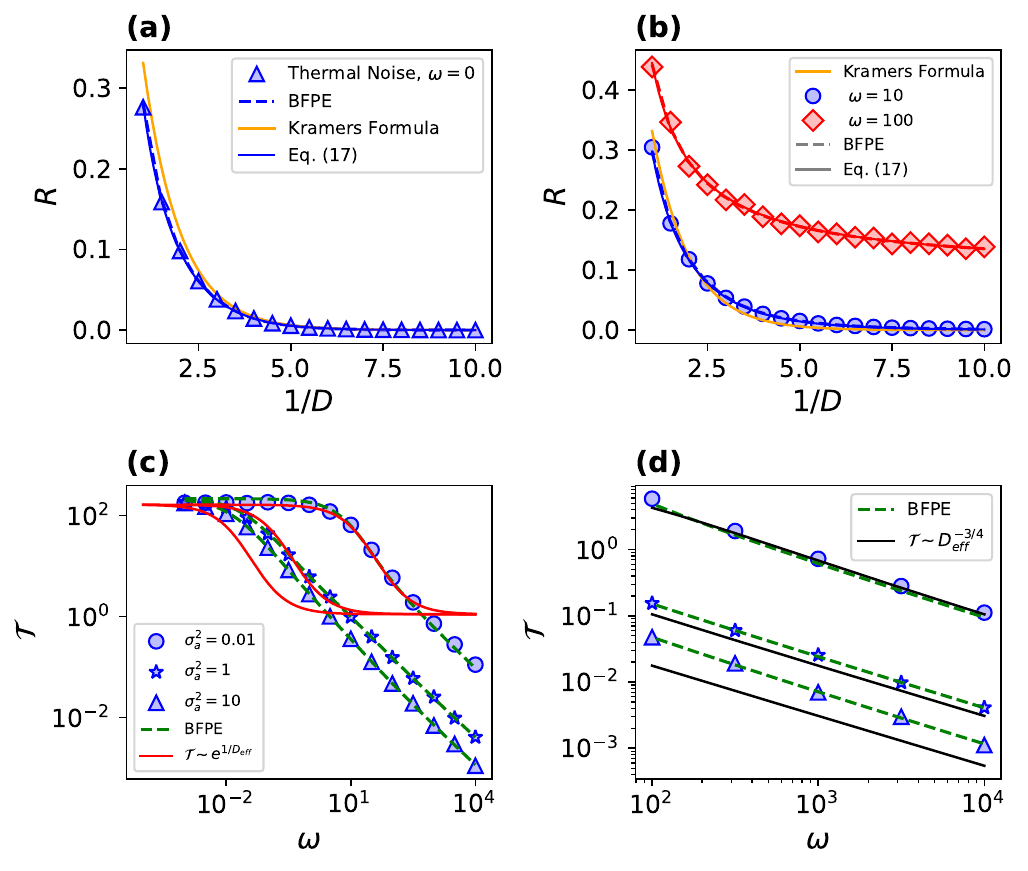}
    \caption{ \textbf{(a)} Escape rate $R$ versus relative barrier height $1/D$ for a particle in the thermal bath. \textbf{(b)} $R$ versus $1/D$ for a particle in the thermal bath and active bath. Solid orange line in (a) and (b) is the plot of Kramers escape rate Eq. (\ref{eq: Dimensionless form of Kramer escape rate for our case}). Other solid lines in sub-figures (a) and (b) are plots of the analytic expression given by Eq. (\ref{eq:the MFPT_Integral_form_3}). \textbf{(c)} MFPT $\mathcal{T}$ versus kick rate $\omega$ for a particle in the thermal bath and active bath for different kick amplitude variance $\sigma_a^2$. Red lines are the plot of Eq. (\ref{eq:the MFPT_Integral_form_small_limit_3}) for the respective simulation parameters. The observed saturation of red plots is consistent with the dependence $e^{1/D_{eff}}$ for large $D_{eff}$, where $D_{eff}$ increases with an increase in $\omega$. \textbf{(d)} $\mathcal{T}$ versus $\omega$ for particle in the thermal bath and active bath in strong noise regime for different kick amplitude variance $\sigma_a^2$. Solid symbols have the same meaning as given in panel $(c)$ legend. The solid black lines correspond to Eq. (\ref{eq:the MFPT_Integral_form_large_limit}) for respective parameters. The solid points are simulation results obtained from Eq. (\ref{eq: dimensionless overdamped langevin PSN}) for respective parameters. The deviation from analytical results from $\sigma_a^2 = 1$ suggest us the breakdown of small variance approximation and thus provides us an estimate of its range of validity. Dashed lines are the simulation plot of BFPE (\ref{eq:the MFPT_detailed}) for the respective parameters. The simulation time step is $10^{-5}$.}
    \label{fig:particle-escape-rate}
\end{figure}

In Fig. \ref{fig:particle-FPT}, we plot the probability distribution for the First Passage Time, for the thermal and active bath with different kick rates. It can be observed that FPT distribution plots show long tails \cite{perez2021time}, and long tails shrink as we increase the variance of the kick amplitude. On a short time scale relative to the mean time, in the thermal, small-variance case, the probability density decreases, whereas it saturates in the large-variance case. The MFPT drops sharply as the kick rate and variance increases, indicating a significant rise in the effective noise strength. Fig. \ref{fig:particle-escape-rate}(a) shows the plot of escape rate against $1/D$ for the thermal case only, i.e., $\omega = 0$. Here, $1/D$ can be understood as the potential barrier height relative to the bath's thermal energy. For the zero kick rate, the particle’s behavior aligns with the Kramers' formula, Eq. (\ref{eq: Dimensionless form of Kramer escape rate for our case}). However, as the kick rate increases, Fig. \ref{fig:particle-escape-rate}(b), the escape rate rises rapidly and saturates at higher values of $1/D$. This saturation occurs earlier for the large kick rates $\omega$ as seen in the Fig. \ref{fig:particle-escape-rate}(b) for $\omega = 100$, where the escape rate becomes nearly constant. These results indicate that PSN injects additional energy into the bath, which depends on the kick rate; hence, saturation is observed when the thermal energy is low or $1/D$ is large compared to the injected energy.

\subsection{The MFPT and the escape rate of a Brownian particle in the presence of thermal bath and active bath}
\label{subsec:particle-escpae-rate}
We now investigate the escape rate and the MFPT using a probabilistic description. At this level, the stochastic dynamics of the Brownian particle can also be equivalently described by the generalized Fokker-Planck equation associated with the underlying Langevin equation. If $P(x,t|x_0,t_0)$ is defined as the probability density of the particle at position $x$ and time $t$, given that it starts from the point $x_0$ at time $t_0$. Then, the forward Fokker-Planck equation for the system corresponding to Eq. (\ref{eq: dimensionless overdamped langevin PSN}) is \cite{hanggi1978derivations, denisov2009generalized}.
\begin{equation}
    \frac{\partial P(x,t|x_0,t_0)}{\partial t} = \mathcal{L}_{\mathrm{FFP}}(x)P(x,t|x_0,t_0),
\end{equation}
where $\mathcal{L}_{\mathrm{FFP}}$ is the Forward Fokker-Planck operator defined as, 
\begin{equation}
  \mathcal{L}_{\mathrm{FFP}}(x)P(x,t|x_0,t_0) = - \frac{\partial(A(x)P)}{\partial x} + D\frac{\partial^2 P}{\partial x^2} + \omega \int_{-\infty}^{\infty} [P(x-y) - P(x)]~\rho_a(y)\, dy,
  \label{eq:forward f-p operator}
\end{equation}
here, $A(x) = -\frac{dU}{dx}$ is the drift term, and $D$ is the diffusion constant. It is also known as the Kolmogorov-Feller equation \cite{fodor2018non, dubkov2016probability}. To find the MFPT, we define the survival probability $S(x_0, t) =  \int_{-\infty}^{x_a} P(x,t|x_0,t_0) ~dx$, here $x_a$ is absorbing boundary or target site in the second well, the MFPT in terms of $S(x_0,t)$ is $\mathcal T(x_0) = \int_0^{\infty} S(x_0,t)~dt$ \cite{gardiner2004handbook},

If the Backward Fokker-Planck operator is $\mathcal{L}_{BFP}$, then the Backward Fokker-Planck equation (BFPE) can be written as,
\begin{equation}
    -\frac{\partial P(x,t|x_0,t_0)}{\partial t_0} = \mathcal{L}_{\mathrm{BFP}}(x_0)P(x,t|x_0,t_0),
    \label{eq: backward fokker planck}
\end{equation}
where $\mathcal{L}_{\mathrm{BFP}}$ is defined as (See Appendix~\ref{BFFO} for the detailed derivation of $\mathcal{L}_{\mathrm{BFP}}$)
\begin{equation}
\begin{aligned}
  \mathcal{L}_{\mathrm{BFP}}(x_0)P(x,t|x_0,t_0) = A(x_0) \frac{\partial P(x,t|x_0,t_0)}{\partial x_0} + D\frac{\partial^2 P(x,t|x_0,t_0)}{\partial x_0^2}\\
 ~+~ \omega \int_{-\infty}^{\infty} [P(x,t|x_0+y,t_0) - P(x,t|x_0,t_0)]\rho_a(y)\,~ dy.
\end{aligned}
  \label{eq:backward f-p operator}
\end{equation}
Integrating Eq. (\ref{eq: backward fokker planck}) on both sides with respect to $x$, it is straightforward to see that the survival probability also satisfies the Backward Fokker-Planck equation. Upon further integrating with respect to $t$, the following relation is obtained for the MFPT as a function of starting position $x_0$. 
\begin{equation}
   A(x_0) \frac{d \mathcal{T}(x_0)}{d x_0} + D\frac{d^2 \mathcal{T}(x_0)}{d x_0^2} + \omega \int_{-\infty}^{\infty} [\mathcal{T}(x_0 + y) - \mathcal{T}(x_0)]\rho_a(y)\, dy = -1.
  \label{eq:the MFPT_detailed}
\end{equation}
After the Taylor expansion of $\mathcal{T}(x_0+y)$ about $x_0$, Eq. (\ref{eq:the MFPT_detailed}) can be rewritten as 
 \begin{equation}
   A(x_0) \frac{d \mathcal{T}(x_0)}{d x_0} + D \frac{d^2 \mathcal{T}(x_0)}{d x_0^2} + \omega\left(\sum_{n=0}^\infty \frac{1}{n!}(\sigma^2/2)^n \frac{d^{2n}T(x_0)}{dx_0^{2n}}- 1\right)\mathcal{T}(x_0)  = -1.
  \label{eq:the MFPT_detailed_II}
\end{equation}
Incidentally, the higher-order derivatives of $\mathcal{T}(x_0)$  are also present in the equation. Now, we use the small variance approximation, i.e., if the variance of kicks $\sigma_a^2 = m_a/\beta_a$ is small, higher-order derivatives of $\mathcal{T}(x_0)$ can be neglected. This approximation gives the following equation:
\begin{equation}
   A(x_0) \frac{d \mathcal{T}(x_0)}{d x_0} + \left(D + \frac{m_a\omega}{2\beta_a}\right)\frac{d^2 \mathcal{T}(x_0)}{d x_0^2}  = -1,
  \label{eq:the MFPT_detailed_2}
\end{equation}
where, let us define the active diffusion $D_a = \frac{1}{2} \sigma_a^2 \omega = \frac{m_a\omega}{2\beta_a}$ and the effective diffusion term as  $D_{eff} = D + D_a$. The active diffusion can be controlled independently by varying the kick variance $\sigma_a$ and the kick rate $\omega$. The active diffusion term $D_a$ is the same as the active temperature of the active bath discussed in the work \cite{di2024brownian}. Thus, PSN under small variance approximation can be treated as thermal noise with active diffusion $D_a$. For studying the behavior in the large variance limit, higher-order derivatives need to be considered in Eq. (\ref{eq:the MFPT_detailed_II}), and Eq. (\ref{eq:the MFPT_detailed}) can be solved numerically to obtain the MFPT for the general case.
The integral form of Eq. (\ref{eq:the MFPT_detailed_2}) is given below \cite{gardiner2004handbook},
\begin{equation}
    \mathcal{T}(x_0) = \frac{1}{D_{eff}}\int_{x_0}^{x_{ab}} dx ~e^{U(x)/D_{eff}} \int_{x_r}^{x}dx'~e^{-U(x')/D_{eff}},
    \label{eq:the MFPT_Integral_form}
\end{equation}
where $x_r$ is the reflecting boundary, $x_0$ is the starting position of the particle, $x_{ab}$ is the target point or absorbing boundary, and $U(x)$ is the potential in which the particle is trapped. In our case, the potential is bistable $U(x) = x^4 - 2x^2$, reflecting boundary $x_r = -\infty$ and particle starts from the left well, and the MFPT is computed when it first arrives in the right well; therefore, $x_0 = -x_m$ and $x_{ab} = x_m$. The expression for the MFPT, given by Eq. (\ref{eq:the MFPT_Integral_form}), can be written in the following form (see appendix \ref{appendix:particle_derivation} for detailed derivation of Eq. (\ref{eq:the MFPT_Integral_form_3}), Eq. (\ref{eq:the MFPT_Integral_form_small_limit_3}), and Eq. (\ref{eq:the MFPT_Integral_form_large_limit})),
\begin{equation}
    \begin{aligned}
    \mathcal{T}(x_0 = -x_m) = \frac{1}{D_{eff}} 2^{-3/2} e^{\frac{1}
    {2D_{eff}}}\left( K_{1/4} \left(\frac{1}{2D_{eff}}\right ) + \pi \sqrt{2}~
    I_{1/4} \left(\frac{1}{2D_{eff}}\right )\right) \\ \int_{-x_m}^{x_m} dx~ e^{(x^4-2x^2)/D_{eff}}.
    \end{aligned}
    \label{eq:the MFPT_Integral_form_3}
\end{equation}
This expression gives the exact MFPT with the diffusion constant $D_{eff}$ for Eq. (\ref{eq:the MFPT_detailed_2}). This central result can now be used to explore different noise regimes and gain analytical understanding in the limiting cases.

In the weak noise regime $(D_{eff} <<1)$, the above MFPT expression can be expressed in the following form,
\begin{equation}
    \mathcal{T}(x_0 = -x_m) \approx \frac{\pi}{2\sqrt{2}} e^{1/D_{eff}}.
    \label{eq:the MFPT_Integral_form_small_limit_3}
\end{equation}
This expression gives the MFPT in the small $D_{eff}$ limit. From the simulation results, we observe that in the weak noise case, $\mathcal{T} >> \tau_{ss}$ where $\tau_{ss} = 1/8$, therefore we can define the escape rate. The inverse of $\mathcal{T}$ gives the escape rate similar to the Kramers' result Eq. (\ref{eq: Dimensionless form of Kramer escape rate for our case}). For the physical insight of the exponential dependence, it is useful to note that escape events are very rare in the weak noise regime, as can be seen in Fig. \ref{fig:particle-trajectory-pdf}(a). Since the probability of crossing the barrier is exponentially suppressed, the escape rate decreases exponentially with decreasing noise strength, leading to an exponential increase in the MFPT.\\
In the strong noise regime $(D_{eff} >>1)$, the MFPT expression Eq. (\ref{eq:the MFPT_Integral_form_3}) can be written as
\begin{equation}
    \mathcal{T}(x_0 = -x_m) \approx  \frac{x_m}{2} \left( \Gamma\left(\frac{1}{4}\right) (D_{eff})^{-3/4} -\frac{1}{2}\Gamma\left(-\frac{1}{4}\right) (D_{eff})^{-5/4}\right).
    \label{eq:the MFPT_Integral_form_large_limit}
\end{equation}

In the strong noise regime, expression of the MFPT consists of two power terms of exponents $-3/4$ and $-5/4$. In the limit of very strong noise, the behavior is dominated by the leading term $D_{eff}^{-3/4}$, while the second term is the sub-leading correction term. For the free process, the MFPT shows power law dependence on $D$ with exponent $-1$. For a bistable potential, in a strong noise regime, the particle is no longer localized near the potential well, but still can not reach the absorbing boundary by means of free diffusion only, since the bistable potential is still influencing the particle. Therefore, in this case, escape is not a rare event, and the exponent $-3/4$ is the artifact of the bistable potential. In this case, we observe from the simulations that $\mathcal{T} \leq \tau_{ss}$, where $\tau_{ss} = 1/8$, we can not define the escape rate in this regime. These results are valid in the regime where the variance of the kicks of active particles is low \cite{pawula1967approximation, risken1989fokker}. For the general case, Eq. (\ref{eq:the MFPT_detailed}) is solved numerically. These results are presented in Figs. \ref{fig:particle-escape-rate}$(c)$ and \ref{fig:particle-escape-rate}$(d)$.


\section{Overdamped Dimer}
\label{sec:dimer}
In many biological and soft-matter processes, the escape rate of multi-monomer units under thermal noise is investigated. Examples include DNA and mRNA translocation through nuclear pores, protein transport across membrane channels \cite{alberts2008molecular,mishra2025competitive}, and polymer translocation through nanopores \cite{pu2016polymer,lubensky1999driven, palyulin2014polymer, sharma2026polymer}. Motivated by these studies, we investigate the escape problem of a minimal composite system, namely a dimer, in the presence of non-Gaussian active noise.

A dimer is shown in Fig. \ref{fig:bistable_potential}(b) in a bistable potential. The interaction between the two particles of the dimer is modeled using two potentials: the harmonic and the Morse potential. The Morse potential does not allow particles to come very close to each other. A harmonic potential with spring constant $k$ is also introduced because anharmonicity of the Morse potential may lead to bond dissociation of the dimer. Harmonic potential is expressed as 
\begin{equation}
    U_{{h}} = \frac{k}{2}(x_2 - x_1-a_0)^2,
    \label{eq:harmonic_potential}
\end{equation}
where $a_0$ is the equilibrium length between the two monomers, $k$ is the spring constant, and $x_2$ and $x_1$ are the positions of the right and left monomers, respectively. The Morse potential is expressed as
\begin{equation}
    U_m = \epsilon_0(1-e^{-\alpha(x_2 - x_1-a_0)})^2,
    \label{eq:morse_potential}
\end{equation}
where $\alpha$ controls the width of the potential, and $\epsilon_0$ represents the dissociation energy. Coupled Overdamped Langevin equations represent the dynamics of individual monomers in a thermal bath and an active bath are given below
\begin{equation}
     \gamma\frac{dx_1(t)}{dt} = k(z -a_0) + x_1(t) - x_1^3(t)+ 2\epsilon_0 \alpha(e^{-\alpha(z-a_0)} - e^{-2\alpha(z-a_0)})+\sqrt{2k_BT\gamma}~\xi_1(t) + \gamma \eta_1(t),
\label{eq:dimer_overdamped_1} \\
\end{equation}
\begin{equation}
    \gamma\frac{dx_2(t)}{dt} = -k(z -a_0) +x_2(t) - x_2^3(t)+ 2\epsilon_0 \alpha(e^{-2\alpha(z-a_0)} - e^{-\alpha(z-a_0)})+\sqrt{2k_BT\gamma}~\xi_2(t) + \gamma \eta_2(t),
\label{eq:dimer_overdamped_2}
\end{equation}
where $z = x_2-x_1$, $\xi_i(t)$ for $i = 1,2$ represent thermal noise, and are independent and identically distributed, i.e., $\langle \xi_1\xi_2 \rangle =\langle \xi_1 \rangle \langle\xi_2 \rangle = 0$. And, $\eta_1(t)$ and $\eta_2(t)$ represent PSN acting on the first and second monomer, respectively, and both are independent of each other. Also, $\langle \xi_i\eta_j \rangle =\langle \xi_i \rangle \langle\eta_j \rangle$ for $i,j = 1,2$. The coupled Langevin equations are made dimensionless by introducing dimensionless parameters $\bar{\alpha} = \alpha/\alpha_c$ where $\alpha_c = 1/x_m$, $\bar{l_0} = l_0/x_m$, $\epsilon_0 = \epsilon/U_B$, and $\bar{k} = k/k_c$, where $k_c = 4U_B/x_s^2$. After dropping the over-bars, the dimensionless coupled Langevin equation can now be written as 
\begin{equation}
     \frac{dx_1(t)}{dt} = 4k(z -a_0) + 4(x_1(t) - x_1^3(t))+ 2\epsilon_0 \alpha(e^{-\alpha(z-a_0)} - e^{-2\alpha(z-a_0)})+\sqrt{2D}~\xi_1(t) + \eta_1(t),
\label{eq:dimensionlses_dimer_overdamped_1_PSN} \\
\end{equation}
\begin{equation}
    \frac{dx_2(t)}{dt} = -4k(z -a_0) +4(x_2(t) - x_2^3(t))+ 2\epsilon_0 \alpha(e^{-2\alpha(z-a_0)} - e^{-\alpha(z-a_0)})+\sqrt{2D}~\xi_2(t) + \eta_2(t).
\label{eq:dimensionless_dimer_overdamped_2_PSN}
\end{equation}
The parameters used in the simulation of Eq. (\ref{eq:dimensionlses_dimer_overdamped_1_PSN}) and Eq. (\ref{eq:dimensionless_dimer_overdamped_2_PSN}) are $m_a = 0.01$, $\beta_a =1 $, $\epsilon = 0.01$ and $\alpha =1 0$ unless otherwise stated.
\begin{figure*}[!htbp]
    \centering
    \includegraphics[width=0.9\linewidth]{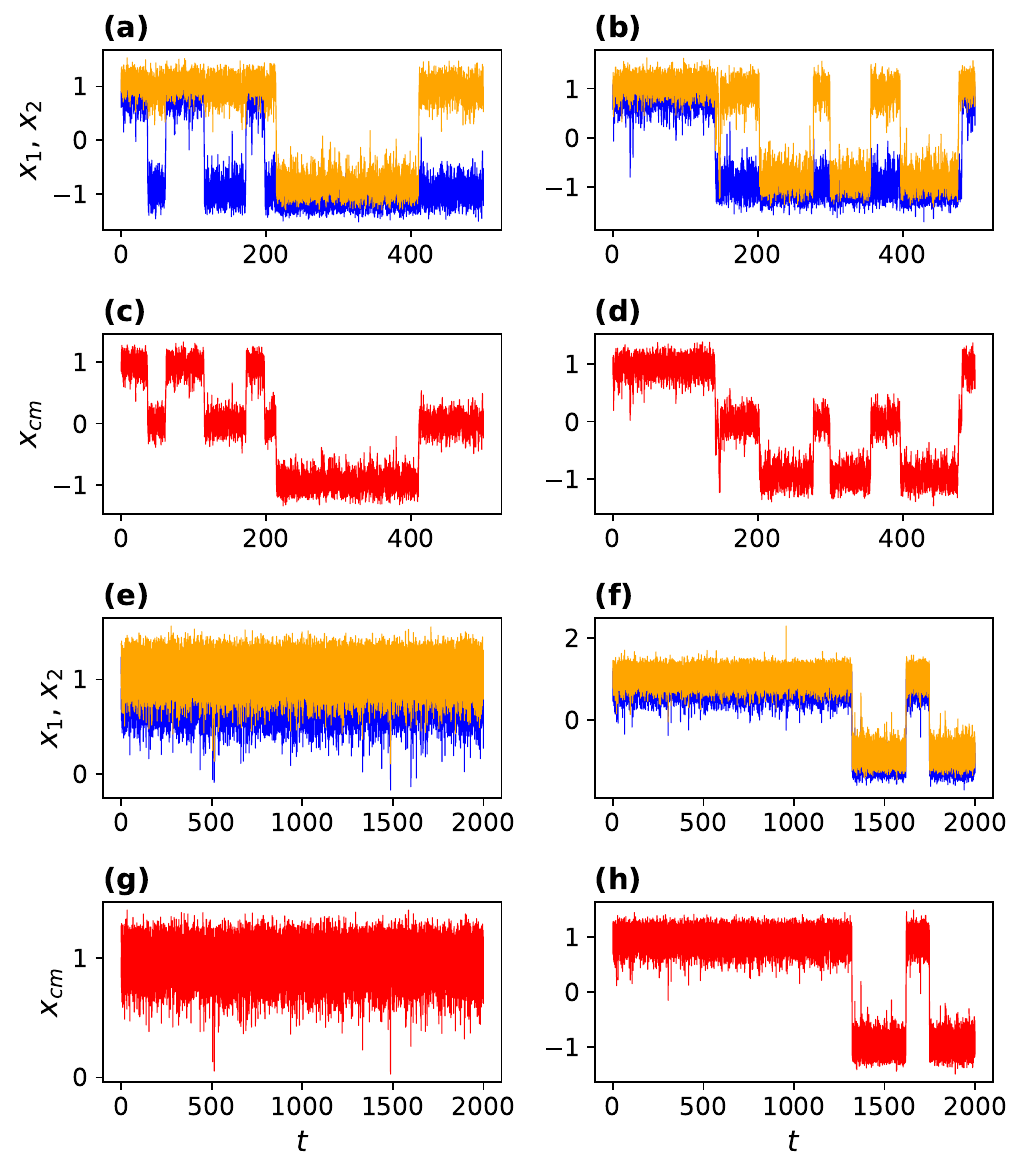}
    \caption{Trajectories of individual monomer positions $x_1, x_2$ (shown in blue and orange colors, respectively), and the center of mass $x_{cm}$ of the dimer as a function of time with $D = 0.2$, subjected to only thermal noise (left panels) and thermal plus Poisson shot noise (active bath) with $\omega = 10$ (right panels). In all simulations, the dimer initially resides in the right well with $x_1(0) = x_2(0) = 1$. The spring constant $k = 0.01$ (first two rows) exhibits a larger number of hopping events, as well as frequent instances when both monomers occupy different wells, compared to $k = 1$ (last two rows), for which hopping events are significantly suppressed, and monomers remain closely bound to each other. The simulation time step is $\Delta t = 10^{-5}$.}
    \label{fig:dimer-trajectory}
\end{figure*}

\begin{figure*}[htbp]
    \centering
    \includegraphics[width=0.9\textwidth]{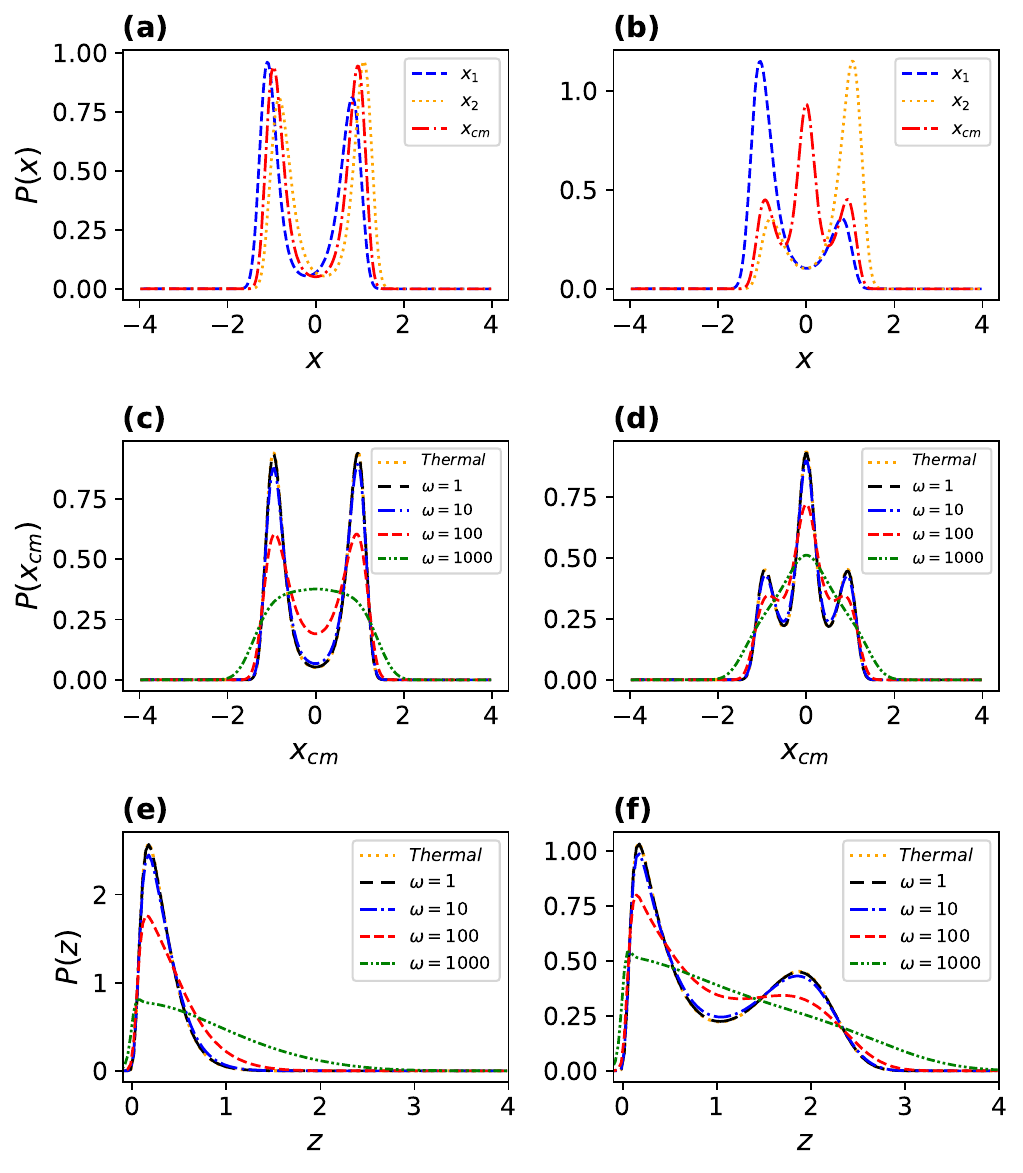}
    \caption{Probability distribution $P(x)$ plot of position $x$ of monomer when dimer is in thermal bath: \textbf{(a)} $k =1$ \textbf{(b)} $k = 0.01$. We observe a large asymmetry in the distribution of $x_1$ and $x_2$ for $k = 0.01$ due to the weak bonding and occurrence of events in which monomers occupy different wells. In contrast, for $k = 1$, both monomers remain closely bound; therefore, asymmetry is small.  $P(x_{cm})$ versus $x_{cm}$ of dimer with different kick rates $\omega$: \textbf{(c)} $k =1$ \textbf{(d)} $k = 0.01$. $P(z)$ versus separation distance $z = x_2-x_1$ of dimer with different kick rates $\omega$: \textbf{(e)} $k =1$ \textbf{(f)} $k = 0.01$. The simulation parameter $D = 0.5$ and time step is $\Delta t = 10^{-5}$.}
    \label{fig:dimer-pdfs}
\end{figure*}
In Fig. \ref{fig:dimer-trajectory}, the trajectories of the individual monomer and the center of mass of the dimer are plotted. The escape events occur more frequently when $k = 0.01$ compared to $k = 1$. For $k = 0.01$, a configuration in which individual monomers residing in different wells is also favorable. Fig. \ref{fig:dimer-pdfs}(a) and Fig. \ref{fig:dimer-pdfs}(b) show the probability distributions for a dimer in contact with a thermal bath. For the center-of-mass coordinate, two peaks are seen for $k = 1$, but for $k = 0.01$, there are three peaks, including one at zero. This additional peak corresponds to the configuration when both monomers reside in different wells. In Fig. \ref{fig:dimer-pdfs}(c) and Fig. \ref{fig:dimer-pdfs}(d), we plot the probability distributions when the dimer is in contact with the thermal bath and active bath for different kick rates. Here, all peaks become shorter and distributions spread as the kick rate increases, and for very large $\omega$, only one peak is seen. In Fig. \ref{fig:dimer-pdfs}(e) and Fig. \ref{fig:dimer-pdfs}(f), the probability distribution plot of the separation distance $z = x_2-x_1$ is shown. $P(z)$ exhibits single peak near equilibrium length $a_0$ for $k =1$ but two peaks one near $a_0$ and other near $z=2$ are present for $k = 0.01.$ Second peak corresponds to the configuration when the particle resides in different wells.

\begin{figure}
    \centering
    \includegraphics[width=\textwidth]{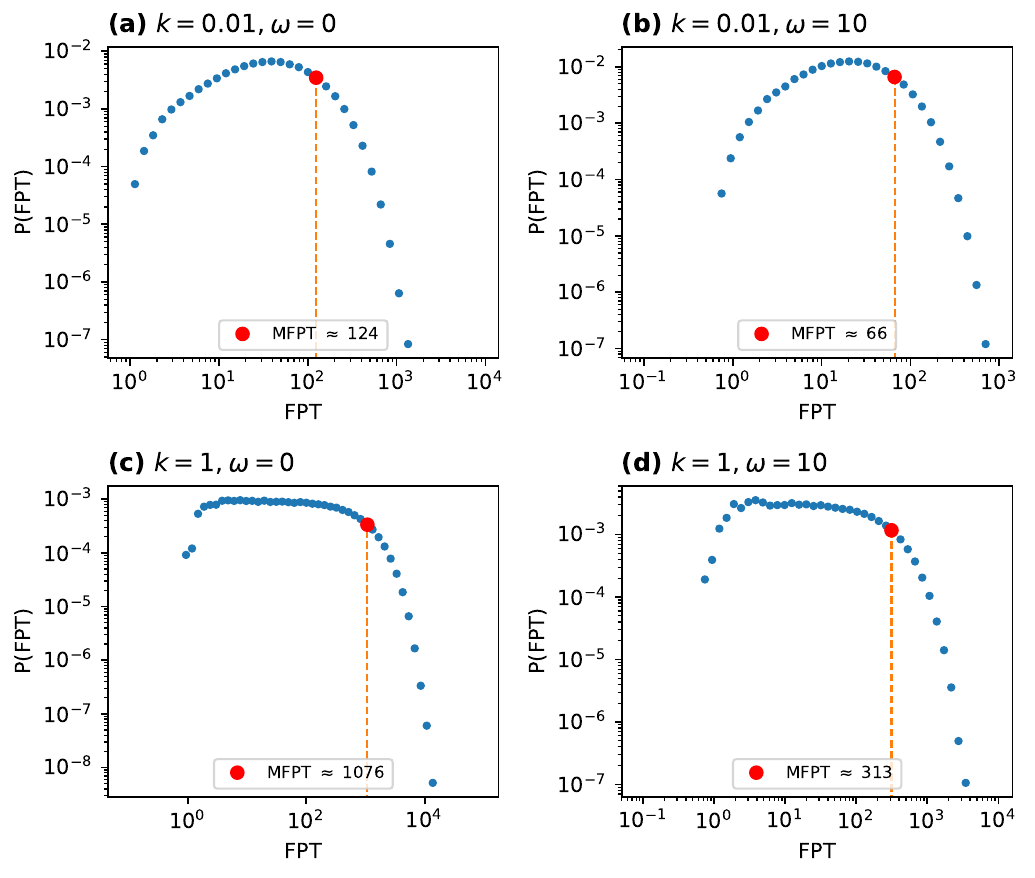}
    \caption{Probability density of the First Passage Time (FPT) for the dimer in thermal bath (left) and thermal plus active bath (right) for kick rate $\omega = 10$. The first and second rows show the FPT distributions for bond strengths $k = 0.01$ and $k = 1$, respectively. Parameter used for the computer simulations is $D=0.2$, $\sigma_a^2 = 0.01,$ and time step $\Delta t = 10^{-5}$. A total of $5\times 10^4$ realizations were used to generate the plot. Long tails in the FPT distributions have similar characteristics as in the single particle case.}
    \label{fig:dimer-fpt}
\end{figure}

The plot of the probability distribution of First Passage Time for the bond strength $k = 0.01$ and $k = 1$ is shown in Fig. \ref{fig:dimer-fpt}. The MFPT turns out to be much lower for $ k=0.01$ than for $ k=1$, suggesting more frequent rare events for $k=0.01$. Using these initial simulation observations, we now try to analytically understand the MFPT of the dimer. First, we only consider the thermal noise, then using results from the single particle, we generalize it to the case when PSN is also present. First, we consider only thermal noise; then, using results from the single-particle case, we generalize to the case where PSN is also present. The MFPT in the small and large bond strength regimes is discussed in the following sections.

\subsection{\texorpdfstring{The MFPT and the escape rate of dimer in small $k$ regime}
{the MFPT and Escape Rate of Dimer in Small k Regime}}
\label{subsec: the MFPT and Escape Rate of Dimer in Small k} 
In the weak bond strength limit, individual monomers can traverse almost independently. Therefore, for the calculation of the MFPT for a dimer, we calculate the MFPT for an individual monomer independently \cite{asfaw2012exploring}. Let the left well be denoted by $-1$ and the right well as $1$. If the left monomer is in the left well and the right monomer is in the right well, the dimer state is then defined as $(-1, 1)$ . In the sequential escape mechanism, we assume that the right monomer first escapes to the right well, followed by the left monomer; however, between the two hop events, the right monomer can return to the left well. This process can be visualized as $(-1, -1) \leftrightarrows (-1, 1) \rightarrow (1, 1)$. Let the escape rate of the right monomer to the right well be $R_1$, the escape rate of the left monomer to the right well be $R_2$, and the escape rate of the right monomer from the right well to the left well be $R_3$. To evaluate the rate, the effective potential experienced by the right monomer when the left monomer is fixed at $x_1 = -1$, can be written as
\begin{equation}
    U_2 = 2k(x_2+1-a_0)^2 + \epsilon_0(1-e^{-\alpha(x_2 +1-a_0)})^2 +x_2^4 - 2x_2^2,
\end{equation}
and, similarly, the effective potential experienced by the left monomer, when the right monomer is fixed at $x_2 = 1$, is written as
\begin{equation}
    U_1 = 2k(1 - x_1-a_0)^2 + \epsilon_0(1-e^{-\alpha(1 -x_1-a_0)})^2 +x_1^4 - 2x_1^2.
\end{equation}
Effective barrier height and curvature at the minimum and the maximum can be evaluated numerically. Following the work \cite{asfaw2012exploring}, the net escape rate from the state $(-1,-1)$ to the final state $(1,1)$ is given by the formula $R = \frac{R_1 R_2}{R_1 + R_2 + R_3}$.

\begin{figure}[!t]
    \centering
    \includegraphics[width=\linewidth]{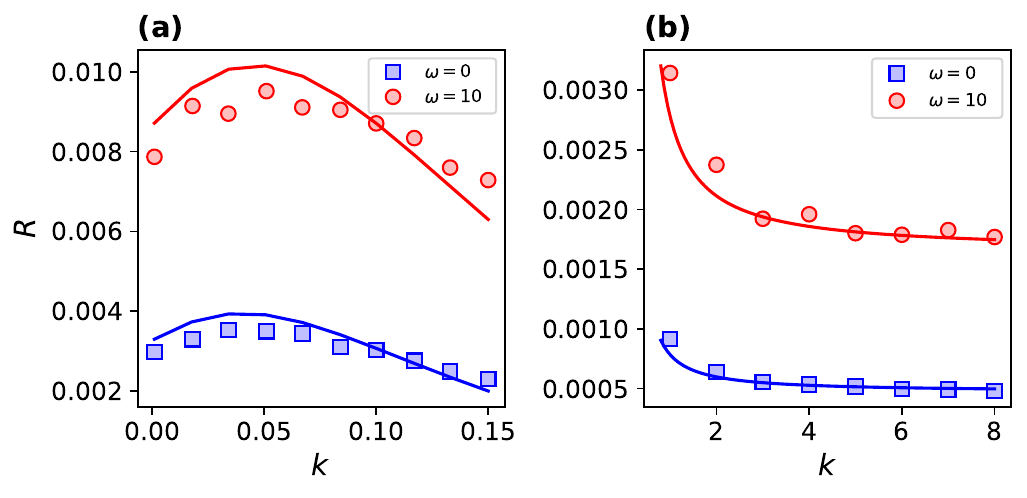}
    \caption{\textbf{(a)} Escape rate $R$ as a function of bond strength $k$ of the dimer. Blue solid squares represent the escape rate for a dimer in a thermal bath, whereas the red solid circles represent the escape rate for the dimer in the thermal bath and active bath. The red and blue solid lines are numerical plots obtained from a sequential-hopping mechanism. \textbf{(b)} Escape rate $R$ as a function of bond strength $k$ of the dimer. Blue solid squares represent the escape rate for a dimer in a thermal bath. whereas the red solid circles represent the escape rate for the dimer in the thermal bath and active bath. The blue and red solid lines are plots for Eq. (\ref{eq: escape rate for the dimer}) with $D$ replaced by $D_{eff}$. It can be observed that the escape rate is not a monotonic function of the bond strength k, and it saturates for large k. The parameters for the computer simulation are $D = 0.2$, $\sigma_a^2 = 0.01$, $\Delta t = 10^{-5}$, and FPT is averaged over $2\times 10^3$ realizations for the computation of the escape rate.}
    \label{fig:dimer-escape-rate-small-large-k}
\end{figure}

As shown in the previous section, in the limit of small kick-variance of PSN, the diffusion term $D$ can be replaced by the effective diffusion term $D_{eff} = D+m_a\omega/2\beta_a$, to approximate the escape rate of the system. Due to Morse potential and harmonic interaction, the effective potential experienced by individual monomers is not symmetric, as it was in the case of the bistable potential; therefore, the results of the previous section cannot be directly used here. However, in the weak noise regime, since the particle follows Kramers formula irrespective of the symmetry of the barrier potential, the expression Eq. (\ref{eq: Dimensionless form of Kramer formula}) can be used to calculate the escape rates of individual monomers. In the strong noise regime, in the case of a single particle, the MFPT follows a power law dependence on the effective noise strength $D_{eff}$. Assuming a similar power law of type $T \sim AD_{eff}^{-3/4} + BD_{eff}^{-5/4}$, data can be fitted to this expression by treating the constants $A$ and $B$ as fitting parameters. In Fig. \ref{fig:dimer-escape-rate-small-large-k}(a), the escape rate of the dimer for thermal noise and thermal plus PSN against $k$ is plotted. For the parameters used in Fig. \ref{fig:dimer-escape-rate-small-large-k}, the MFPT is very large compared to $\tau_{ss}$ therefore we can define escape rate. The solid lines show that the sequential-hopping mechanism approximates the escape rate to a reasonable degree. The escape rate is also not monotonic as a function of the bond strength $k$. Thus, a specific bond strength can optimize the escape rate.  We now consider the other limiting case, namely the large $k$ limit.

\subsection{\texorpdfstring{The MFPT and the escape rate of dimer in large $k$ regime}
{the MFPT and Escape Rate of Dimer in Large k Regime}}
\label{sec:dimer-escape-large-k}
In the large bond strength limit, the relative distance of the dimer will fluctuate and relax very fast compared to the center of mass coordinate of the dimer. In this case, we use the adiabatic
method in which we eliminate the fast variable (relative distance) from our model and look
for the dynamics of the slow variable (center of mass coordinate). Let the center of mass coordinate of the dimer be $x_{cm} = \frac{x_1 + x_2}{2}$ and relative separation distance $r = \frac{x_2 - x_1}{2}$, for now we focus on the thermal bath only and will include PSN effects later, using these substitution, Eq. (\ref{eq:dimensionlses_dimer_overdamped_1_PSN}) and Eq. (\ref{eq:dimensionless_dimer_overdamped_2_PSN}) becomes
\begin{equation}
     \frac{dx_{cm}(t)}{dt} =  4x_{cm}(t) - 4x_{cm}^3(t) - 12x_{cm}r^2+ \frac{\sqrt{2D}}{2}~(\xi_1(t)+\xi_2(t)),
\label{eq:dimensionlses_dimer_com_overdamped_1_1} \\
\end{equation}
\begin{equation}
    \frac{dr(t)}{dt} = -4k(2r -a_0) +4(r - r^3 - 3x_{cm}^2r)+ 2\epsilon_0 \alpha(e^{-2\alpha(2r-a_0)} - e^{-\alpha(2r-a_0)})+\frac{\sqrt{2D}}{2}~(\xi_2(t)-\xi_1(t)).
\label{eq:dimensionless_dimer_com_overdamped_2_2}
\end{equation}
As the sum and difference of two independent and identically distributed Gaussian variables is also a Gaussian, with total variance twice the original variance. If $\xi_{cm} = \frac{\xi_1+\xi_2}{\sqrt{2}}$ and $\xi_r = \frac{\xi_2 -\xi_1}{\sqrt{2}}$ then they satisfy $\langle \xi_{cm}\rangle = \langle\xi_r\rangle = 0$, $\langle \xi_{cm}(t) \xi_{cm}(t')\rangle = \langle\xi_r(t)\xi_r(t')\rangle = \delta(t-t')$, and $\langle \xi_{cm}\xi_r\rangle = 0$. Thus Eqs. (\ref{eq:dimensionlses_dimer_com_overdamped_1_1}) and (\ref{eq:dimensionless_dimer_com_overdamped_2_2}) can be written as
\begin{equation}
     \frac{dx_{cm}(t)}{dt} =  4x_{cm}(t) - 4x_{cm}^3(t) - 12x_{cm}r^2+ \sqrt{D}~\xi_{cm}(t),
\label{eq:dimensionlses_dimer_com} \\
\end{equation}
\begin{equation}
    \frac{dr(t)}{dt} = -4k(2r -a_0) +4(r - r^3 - 3x_{cm}^2r)+ 2\epsilon_0 \alpha(e^{-2\alpha(2r-a_0)} - e^{-\alpha(2r-a_0)})+\sqrt{D}~\xi_{r}(t).
\label{eq:dimensionless_dimer_relative_distance}
\end{equation}
When the bond strength $k$ becomes very large, the restoring force acting on the relative separation distance of the dimer increases correspondingly. As a result, for the stiff bonds, the coordinate $r$ fluctuates rapidly in a small region around the equilibrium position, leading to a fast relaxation of fluctuations in the coordinate $r$. In contrast, the motion of the center of mass of the dimer is relatively slow compared to the motion of the relative separation distance of the dimer. This clear separation of time scales allows us to treat $r$ as a fast variable. Thus, by eliminating $r$, an effective description for the dynamics of the center-of-mass can be obtained. This procedure, known as adiabatic elimination, was initially developed in quantum mechanics and has since been widely applied to classical stochastic systems, see for example, \cite{kaneko1981adiabatic,risken1989fokker,lan2008elimination, asfaw2012exploring} for a review on this method and its applications. Our goal is to find out the expectation value of the drift term in Eq. (\ref{eq:dimensionlses_dimer_com}) in the stationary state of the relative coordinate $r$. To find the stationary distribution of the relative coordinate $r$, we write the Fokker-Planck equation corresponding to the Langevin equation (\ref{eq:dimensionless_dimer_relative_distance}). Let $\Phi(r)$ denote the stationary distribution of coordinate $r$, the governing Fokker-Planck equation can be written as,
\begin{equation}
    \frac{d}{dr}\left[B\Phi(r) \right] + \frac{D}{2}\frac{d^2\Phi}{dr^2} = 0,
    \label{eq: fast variable distribution}
\end{equation}
where, $B = U' +V'$ where $ U' = 4\kappa(2r-a_0)- 2\epsilon_0\alpha(e^{-2\alpha(2r-a_0)} -e^{-\alpha(2r-a_0)})$ and $V' = -4(r-r^3-3x_{cm}^2r)$. The harmonic and Morse terms in $U'$ force the relative coordinate to fluctuate about the equilibrium position $r^* = a_0/2$, but there is an additional term $V'$ due to the bistable potential and a cross term of fast and slow variables, which shifts the equilibrium position from $a_0/2$ to some new position. Let this new equilibrium position be $r_0 = r^* + \delta r$. The first order correction in the equilibrium position can be calculated by the perturbative approach. Since, $\delta r = -V'(r^*)/U''(r^*)$, it becomes $\delta r= \frac{4(r^* -r^*{^3} -3x_{cm}^2r^*)}{8k+4\epsilon_0\alpha^2}$. For $r^* = a_0/2$, this gives $\delta r= \frac{4a_0 -a_0^3 -12x_{cm}^2a_0}{8(2k+\epsilon_0\alpha^2)}$.  Now, $B(r)$ can be Taylor expanded about the equilibrium position $r_0$.
\begin{equation}
    B(r) = B(r_0) + \left. \frac{dB}{dr} \right|_{r = r_0} (r-r_0)+...,
\end{equation}
where $B(r_0) = 0$ and $\left. \frac{dB}{dr} \right|_{r = r_0} = 8k-4+3a_0^2 + 12x_{cm}^2+4\epsilon_0\alpha^2 + O(\delta r) = F$. Therefore, $B(r)$ can be written as $B(r) = F(r-r_0)$. Integrating Eq. (\ref{eq: fast variable distribution}) and using the fact that $\Phi$ should vanish as $r \rightarrow \infty$, gives
\begin{equation}
    \Phi(r) = \mathcal{N}e^{-\frac{(r-r_0)^2}{D/F}}.
\end{equation}
This is a Gaussian distribution with mean $r_0$ and variance $D/2F = D/16k + O(1/k^2) $. $\mathcal{N}$ is the normalization constant. Using this, drift of the slow variable averaged over the stationary distribution $\Phi(r)$ can be calculated. The effective center of mass Langevin equation reads
\begin{equation}
     \frac{dx_{cm}(t)}{dt} =  4x_{cm}(t) - 4x_{cm}^3(t) - 12x_{cm} \langle r^2 \ran+ \sqrt{D}~\xi_{cm}(t),
\label{eq:dimensionlses_dimer_com_1} \\
\end{equation}
and $\langle r^2 \ran = r_0^2 +\frac{D}{16k}$. Since, $r_0 = a_0/2 + \delta r$, this gives, $r_0^2 = a_0^2/4 + a_0 \delta r + O((\delta r)^2)$ where $a_0 \delta r= \frac{4a_0^2 -a_0^4}{8(2k+\epsilon_0\alpha^2)} - \frac{12x_{cm}^2a_0^2}{8(2k+\epsilon_0\alpha^2)} $. 
Substituting these values, Eq. (\ref{eq:dimensionlses_dimer_com_1}) becomes
\begin{equation}
     \frac{dx_{cm}(t)}{dt} =  ax_{cm}(t) - bx_{cm}^3(t) + \sqrt{D}~\xi_{cm}(t),
\label{eq:dimensionlses_dimer_com_2} \\
\end{equation}
where $a = 4 - 3a_0^2- \frac{3(4a_0^2 -a_0^4)}{2(2k+\epsilon_0 \alpha^2)} - \frac{12D}{16k}$ and $ b = 4- \frac{18a_0^2}{2k + \epsilon_0\alpha^2}$. The Langevin equation (\ref{eq:dimensionlses_dimer_com_2}) gives the dynamics of a particle trapped in a bistable potential of the form $U = \frac{-ax_{cm}^2}{2} + \frac{bx_{cm}^4}{4}$ with barrier at $x_{cm} = 0$ of height $a^2/4b$ in contact with a thermal bath with noise strength $D/2$. Therefore, following the results of the previous section, the MFPT for a dimer in a large $k$ regime is
\begin{equation}
    \mathcal{T}(x_0 = -x_m) = \frac{2}{D} 2^{-3/2} \sqrt{\frac{a}{b}}e^{\frac{a^2}{4bD}}\left( K_{1/4} \left(\frac{a^2}{4bD}\right ) + \pi \sqrt{2} I_{1/4} \left(\frac{a^2}{4bD}\right )\right) \int_{-x_m}^{x_m} dx ~e^{(\frac{b}{4}x^4-\frac{a}{2}x^2)/\frac{D}{2}}.
    \label{eq:the MFPT_Integral_form_dimer_com}
\end{equation}
In the limit of weak noise i.e. $\frac{D}{2}<<1$, the escape rate that is the inverse of the MFPT exhibits Arrhenius scaling
\begin{equation}
    R = \frac{\sqrt{2}a}{2\pi} e^{\frac{-a^2}{2bD}}.
    \label{eq: escape rate for the dimer}
\end{equation}
This expression for escape rate is consistent with the Kramers result given in Eq. (\ref{eq: Kramer Escape Rate formula}). In the strong noise limit i.e. $\frac{D}{2}>>1$, the MFPT Eq. (\ref{eq:the MFPT_Integral_form_dimer_com}) reduces to the following power law expression
\begin{equation}
    \mathcal{T}(x_0 = -x_m) \approx 2x_m 2^{-3/2} \sqrt{\frac{a}{b}} \left(  \left(\frac{a^2}{8b}\right )^{-1/4}\Gamma\left(\frac{1}{4}\right) D^{-3/4} - \left(\frac{a^2}{8b}\right )^{1/4}\Gamma\left(-\frac{1}{4}\right) D^{-5/4}\right).
    \label{eq:the MFPT_Integral_form_large_limit_diimer}
\end{equation}
Similar to the particle escape, dimer also shows a power law dependence on $D$ in the strong bond regime. In the limit $k \rightarrow \infty$ and $a_0 \rightarrow 0$, the constants are $a = b = 4$, and center of mass coordinate Eq. (\ref{eq:dimensionlses_dimer_com_2}) reduces to the following equation, which represents the dynamics of a globular particle (composite of two monomers) with diffusion constant $D/2$,
\begin{equation}
     \frac{dx_{cm}(t)}{dt} =  4(x_{cm}(t) - x_{cm}^3(t) )+ \sqrt{D}~\xi_{cm}(t).
\label{eq:dimensionlses_dimer_com_limit_overdamped_1} \\
\end{equation}
And, the escape rate of this system in the weak noise limit is
\begin{equation}
R = \frac{2\sqrt{2}}{\pi} e^{-2/D}.
\label{eq: R with D}
\end{equation}
To note that according to section \ref{sec: A single Brownian particle}, in the presence of PSN, under the small variance approximation, the diffusion coefficient can be replaced by the effective diffusion coefficient $D_{eff} = D + m_a\omega/2\beta_a$ in Eqs. (\ref{eq:the MFPT_Integral_form_dimer_com}), (\ref{eq: escape rate for the dimer}), (\ref{eq:the MFPT_Integral_form_large_limit_diimer}) and (\ref{eq: R with D}).
\begin{figure}
    \centering
    \includegraphics[width=\linewidth]{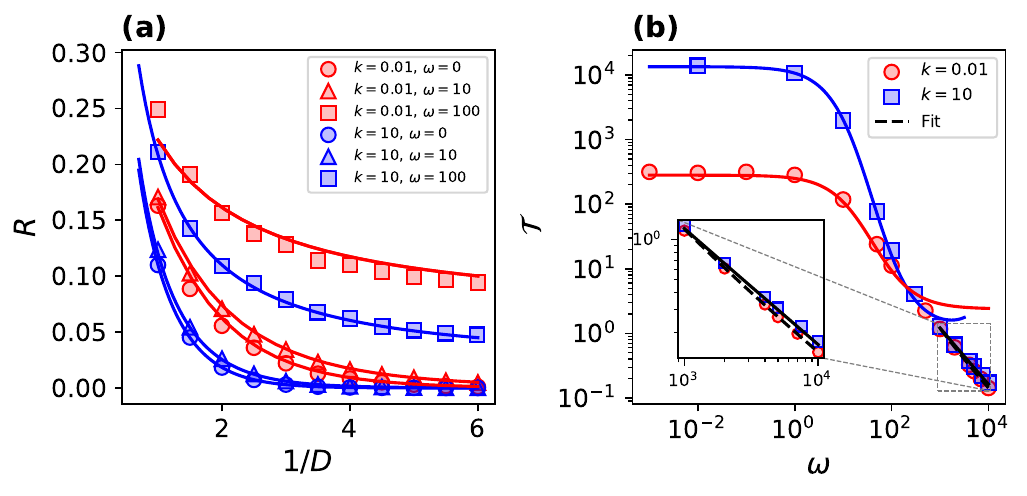}
    \caption{\textbf{(a)} Escape rate $R$ as a function of $1/D$ for the dimer for different bond strength $k$. The red solid lines are plots of the escape rate obtained using the sequential-hopping mechanism for $k = 0.01$ as discussed in section \ref{subsec: the MFPT and Escape Rate of Dimer in Small k}. The blue solid lines are the plots of Eq. (\ref{eq:the MFPT_Integral_form_dimer_com}) with $k=10$ for respective parameters. \textbf{(b)} the MFPT as a function of kick rate $\omega$ is plotted. Red solid line plot is obtained for $k = 0.01$ using the sequential hopping mechanism. The blue solid line is the plot of Eq. (\ref{eq: escape rate for the dimer}) for $k = 10$ with $D$ replaced by $D_{eff}$. The black solid line is the plot of Eq. (\ref{eq:the MFPT_Integral_form_large_limit_diimer}) for the strong noise case for $k = 10$. The black dashed line is a fit plot for $k = 0.01$ with power law ${\mathcal T} \sim AD_{eff}^{-3/4}+BD_{eff}^{-5/4}$. Both axes are in log scale. The inset figure shows the rightmost data points corresponding to high kick rates. The solid symbol obtained from the computer simulation of Eq. (\ref{eq:dimensionlses_dimer_overdamped_1_PSN}) and (\ref{eq:dimensionless_dimer_overdamped_2_PSN}). The parameters for the computer simulation are  $D = 0.2$, $\sigma_a^2 = 0.01$, time step is $\Delta t = 10^{-5}$ and decreased for large $\omega$ values, and FPT is averaged over $2\times 10^3$ realizations.}
    \label{fig:dimer-escape-rate}
\end{figure}
Fig. \ref{fig:dimer-escape-rate-small-large-k}(b) shows the comparison of the analytical results with the simulations in the large $k$ limit. The escape rate against bond strength $k$ is plotted. The blue solid line plot shows only the thermal noise, and the red solid line plot shows the dimer in simultaneous contact with the thermal and PSN. We see that as the bond strength $k$ increases, the MFPT converges to the value given by Eq. (\ref{eq: R with D}).

Fig. \ref{fig:dimer-escape-rate} summarizes all the findings for the dimer case. Fig. \ref{fig:dimer-escape-rate}(a) shows the behavior of the escape rate as a function of $ 1/D$. Fig. \ref{fig:dimer-escape-rate}(b) shows the MFPT as a function of the kick rate $\omega$. In the weak noise case, simulation data agree with the sequential hopping mechanism for $k = 0.01$ and Eq. (\ref{eq: escape rate for the dimer}) for $k = 10$. In the strong noise case, simulation verifies that the dimer with strong bond strength follows the power law given by Eq. (\ref{eq:the MFPT_Integral_form_large_limit_diimer}). Similar to the stiff dimer and the particle case in strong noise, the MFPT is fitted with a power law for weak bond strength.

\section{Effect of an intermediate barrier on the MFPT}
\label{sec:intermediate-potential}
 A  recent theoretical and experimental study \cite{chupeau2020optimizing} showed that the escape rate of a particle is not determined solely by the height of the barrier, but the shape of the potential profile can significantly modify it. This study shows that an optimized N-shaped profile accelerates the exit dynamics, reducing the MFPT by a factor of two relative to free diffusion. Recent work \cite{kumar2025speeding} also investigated speed-ups achieved through the introduction of an intermediate barrier or kink in linear and harmonic potential profiles with finite boundary in the presence of delta correlated white noise. Motivated by these studies, we also investigate the effect of an intermediate barrier on the MFPT of a particle and a dimer in contact with a thermal bath and an active bath.

\begin{figure}[!ht]
    \centering
    \includegraphics[width=0.5\linewidth]{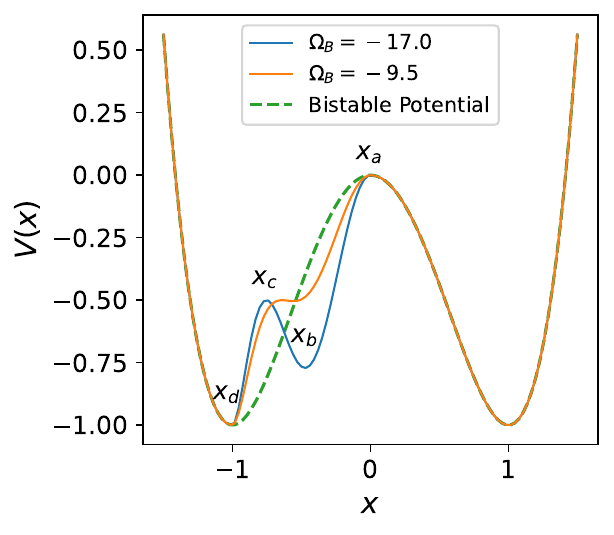}
    \caption{Piecewise Potential (\ref{eq:piecewise_potential}) for intermediate barrier height $V(x_c) = -0.5$ for curvature values $\Omega_B = -17$ (Blue plot) and $\Omega_B = -9.5$ (Orange plot). For comparison, the dashed plot corresponds to the bistable potential. The points $x_a,~x_b,~x_c,~x_d$ can be adjusted to produce different potential profiles.}
    \label{fig:piecewise_potential}
\end{figure}
\begin{figure}[!htbp]
    \centering
\includegraphics[width=\linewidth]{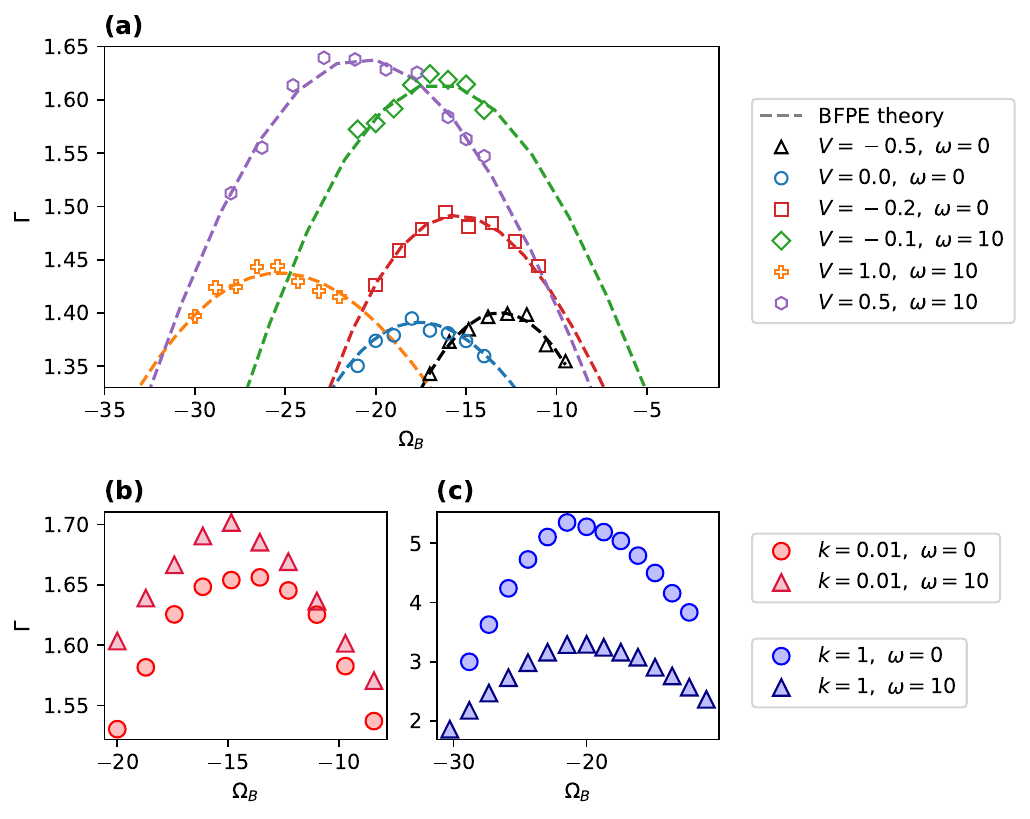}
    \caption{\textbf{(a)} Relative MFPT $\Gamma$ of Brownian particle as a function of curvature $\Omega_B$ for different intermediate potential heights. The dashed lines are the numerical plot of Eq. (\ref{eq:the MFPT_detailed}) for the piecewise potential (\ref{eq:piecewise_potential}). Solid symbols indicate simulation results for corresponding dashed lines. We observe that a suitable intermediate barrier reduces the MFPT. \textbf{(b)} Relative escape rate $\Gamma$ of Dimer as a function of curvature $\Omega_B$ for intermediate potential height $V = -0.2$ for bond strength $k = 0.01$ and \textbf{(c)} $k = 1$. Solid circles correspond to the rate for thermal noise only, whereas solid triangles correspond to the rate when both thermal and PSN noise are present. The parameters used for the simulation are $D = 0.25, m_a = 0.01, \beta_a = 10$, $\sigma_a^2 = 0.01$, $\Delta t = 10^{-5}$ with an ensemble average of $5 \times10^{5}$. }
    \label{fig: relative escape rate for particle and dimer}
\end{figure}
A barrier is introduced between the maximum at $x_a = 0$ and the minimum at $x_d = -1$ of the bistable potential. Because of this, the original bistable potential is modified, as shown in Fig. \ref{fig:piecewise_potential}. The functional form of this piecewise potential profile can be written as 
\begin{equation}
V(x)=
\begin{cases}
\displaystyle
\frac{-\Omega_B}{x_b^{2}x_c^{2}x_d^{2}}
\Bigg[
\frac{x^{8}}{8}
-\frac{x^{6}}{6}
\left(x_b^{2}+x_c^{2}+x_d^{2}\right)
+\frac{x^{4}}{4}
\left(
x_b^{2}x_c^{2}
+x_b^{2}x_d^{2}
+x_c^{2}x_d^{2}
\right)
\\[8pt]
\displaystyle \qquad
-\frac{x^{2}}{2}
x_b^{2}x_c^{2}x_d^{2}
\Bigg],
& x_d \le x \le x_a,
\\[12pt]
x^{4}-2x^{2},
& \text{otherwise},
\end{cases}
\label{eq:piecewise_potential}
\end{equation}
where $x_a, x_b, x_c$ and $ x_d$ are the stationary points of $V(x)$ ordered from right to left $(0 = x_a >x_b>x_c>x_d = -1)$. And, $\Omega_B$ is the curvature of an eighth-degree polynomial at $x_a$. The parameters are chosen such that the potential is $ -1$ at $x_d =-1$. The $x_b$ and $x_c$ control the height of the intermediate barrier at $x_c$ for a given $\Omega_B$. Let the relative MFPT parameter $\Gamma$ be
\begin{equation}
    \Gamma = \frac{\mathcal{T}_0}{\mathcal{T}(\Omega_B, V(x_c))},
\end{equation}
where $\mathcal{T}_0$ is the MFPT of the Brownian particle in the bistable potential, and  $\mathcal{T}(\Omega_B, V(x_c))$ is the MFPT for the modified piecewise potential. If $\Gamma >1$, this indicates that the average time to cross the barrier decreases. Fig. \ref{fig: relative escape rate for particle and dimer} (a) shows the plot of the relative MFPT for the Brownian particle for different intermediate potential heights and curvatures $\Omega_B$. The figure indicates that the relative MFPT attains an optimal value for intermediate barrier heights at a specific curvature. Relative MFPT converges towards zero for very large $\Omega_B$ as expected. It should also be noted that the optimal intermediate height for the passive system is approximately $V(x_c) = -0.2$. In contrast, when the Brownian particle is simultaneously in contact with the thermal bath and active bath, the optimal intermediate barrier height becomes approximately $V(x_c) = 0.5$. Simulations of Eq. (\ref{eq:the MFPT_detailed}) for the potential (\ref{eq:piecewise_potential}) also validate these results. For the large noise strength, the effect of potential on the particle will be minimal, and the crossing events will not be rare; therefore, this case will not help to speed up the escape process.

Figures \ref{fig: relative escape rate for particle and dimer} (b) and (c) show the simulation results for the MFPT of the dimer in the presence of an intermediate potential. For an intermediate barrier height of $V = -0.2$, the weakly bonded dimer and the stiff dimer's MFPT decrease in both thermal and thermal plus PSN cases. For these parameters, the relative MFPT for the thermal noise is higher compared to the case when both thermal and PSN are present for $k =1$, while the opposite trend is seen for $k = 0.01$.

\section{Conclusion And Discussion}
\label{conclude}
In this work, we have investigated the MFPT of a Brownian particle and a dimer in the presence of thermal and active noises, in a bistable potential, as well as in the presence of intermediate barriers. Active noise was modeled to follow Poisson statistics. First, the MFPT is formulated using the small-variance approximation. We showed that the escape rate for the particle exhibits Kramers' formula in the weak noise regime with effective noise strength $D_{eff}$, and the MFPT showed a power law dependence on $D_{eff}$ in the strong noise regime. Under the small-variance approximation, the MFPT of a dimer is also investigated in the weak and strong bond strength regimes. The MFPT of the dimer was found to depend strongly on the bond strength. A sequential hopping mechanism was used for a small bond-strength dimer, and numerical results matched with simulations. Additionally, the adiabatic elimination method was used in the large bond strength regime to formulate the MFPT of the dimer's center of mass. In this case, the dimer follows Arrhenius scaling for escape rate in the weak-noise regime and a power law for the MFPT in the strong-noise regime, similar to the single-particle case but with half the diffusion constant. The MFPT of the Brownian particle and also of the dimer were studied in the presence of an intermediate barrier. For certain intermediate barrier profiles, the MFPT decreases. This reduction in the MFPT can be optimized by tuning the curvature $\Omega_B$ and the intermediate barrier height. Interestingly, it was observed that for a particle in contact with a thermal and an active bath, an intermediate barrier higher than the original barrier yields an optimal escape event. In the limiting case of small variance, analytical calculations match very well with the numerical and simulation results. For the general case, when higher-order terms are present in Eq. (\ref{eq:the MFPT_detailed_II}), we can not compare the active bath with a thermal bath in terms of effective diffusion coefficients alone. But we can observe from Fig. \ref{fig:particle-escape-rate}, panel (d), for variance $\sigma_a^2 = 1$ and $\sigma_a^2 = 10$, that the analytical result from the truncated series underestimates the MFPT, but the slope therefore exponent retains its value for the MFPT. Representative results for the MFPT when kick amplitudes follow distributions other than the Gaussian, namely, bimodal, uniform, and Laplace, are also provided in the appendix \ref{nongauss}. In the case of large variance, the MFPT displays a strong dependence on the specific form of kick distribution. Under identical variance conditions, the MFPT is minimum for a bimodal distribution and increases monotonically for uniform, Gaussian, and Laplace. In the case when variance of all distributions is very small, all four distributions yield similar MFPT.

Our work provides insight into the MFPT of a passive Brownian particle and a dimer in a thermal bath and an active bath, trapped in a bistable potential, with and without an additional intermediate barrier. An analytical description of the MFPT in the strong kick-variance limit remains an open question. Some recent theoretical and experimental studies have examined the crossing time and transition path time of a particle with a memory kernel \cite{ferrer2024transition, netz2026barrier}. This suggests a natural extension of this work to the non-Markovian inertial systems, viscoelastic baths, and correlated Poisson shot noise. Moreover, recent analytical studies on two-dimensional active particles and their first passage time distributions motivate a natural generalization of the present work to higher dimensions \cite{caraglio2022analytic, caraglio2025two}. It would also be interesting to generalize the study to the escape dynamics of a harmonic chain in the presence of active noise.  

\appendix
\counterwithin{figure}{section}
\counterwithin{table}{section}

\section{Backward Fokker-Planck Operator}
\label{BFFO}
This appendix gives details of the calculation for the backward Fokker-Planck equation that we use in Section \ref{subsec:particle-escpae-rate}. The Forward Fokker-Planck Equation for the system following equation (\ref{eq: dimensionless overdamped langevin PSN}) is \cite{hanggi1978derivations, denisov2009generalized}.
\begin{equation}
    \frac{\partial P(x,t)}{\partial t} = \mathcal{L}_{\mathrm{FFP}}P(x,t).
\end{equation}
We can define the Forward Fokker-Planck operator $\mathcal{L}_{\mathrm{FFP}}$ as
\begin{equation}
  \mathcal{L}_{\mathrm{FFP}}P(x,t) = - \frac{\partial(A(x)P(x,t))}{\partial x} +~D \frac{\partial^2 P(x,t)}{\partial x^2} + \omega \int_{-\infty}^{\infty} [P(x-y,t) - P(x,t)]\rho_a(y)\, dy.
  \label{eq:forward f-p operator_append}
\end{equation}
Let the Backward Fokker-Planck operator be $\mathcal{L}_{\mathrm{BFP}}$, and since both operators form an adjoint pair, the two functions $\phi$ and $P$ follow
\begin{equation}
    \langle \phi, \mathcal{L}_{\mathrm{FFP}} P(x,t) \rangle  = \langle \mathcal{L}_{\mathrm{BFP}} \phi, P(x,t) \rangle,
\end{equation}
or, in integral form
\begin{equation}
    \int_{-\infty}^{\infty} \phi (\mathcal{L}_{\mathrm{FFP}} P(x,t) )\,dx   = \int_{-\infty}^{\infty} P(x,t) (\mathcal{L}_{\mathrm{BFP}} \phi)\,dx.
    \label{adjoint relation}
\end{equation}
Drift Term:
\begin{equation}
    \int_{-\infty}^\infty \phi \frac{d}{dx}(AP(x,t))\,dx = \phi A P(x,t) |_{-\infty}^{\infty} -  \int_{-\infty}^\infty \left(\frac{d\phi}{dx}\right) AP(x,t) \,dx.
\end{equation}
Assuming $P(x,t) \to 0 \quad \text{as} \quad x \to \pm\infty$, this gives
\begin{equation}
    \int_{-\infty}^\infty \phi \frac{d}{dx}(AP(x,t))\,dx =  -  \int_{-\infty}^\infty P(x,t) A \left(\frac{d\phi}{dx}\right)  \,dx .
    \label{Drift term in adjoint relation}
\end{equation}
Diffusion Term:
\begin{equation}
     \int_{-\infty}^\infty \phi \frac{d^2}{dx^2}\left(AP(x,t)\right)  = \phi D \left.\frac{dP(x,t)}{dx}\right|_{-\infty}^{\infty} -  \int_{-\infty}^\infty\frac{d\phi}{dx} D\frac{dP(x,t)}{dx}\, dx,
\end{equation}
\begin{equation}
     \int_{-\infty}^\infty \phi \frac{d^2}{dx^2}\left(AP(x,t)\right)  = -\left.\frac{d\phi}{dx} DP(x,t)\right|_{-\infty}^{\infty} +  \int_{-\infty}^\infty  DP(x,t) \frac{d^2\phi}{dx^2} \, dx,
\end{equation}
\begin{equation}
     \int_{-\infty}^\infty \phi \frac{d^2}{dx^2}\left(AP(x,t)\right)  =   \int_{-\infty}^\infty  DP(x,t) \frac{d^2\phi}{dx^2} \, dx.
     \label{Diffusion term in adjoint relation}
\end{equation}
PSN term:
\begin{equation}
\begin{aligned}
\int_{-\infty}^{\infty} 
\phi\!\left( \omega \int_{-\infty}^{\infty} [P(x-y,t) - P(x,t)] \rho_a(y)\, dy \right) dx 
&= 
\int_{-\infty}^{\infty} 
\phi\!\left( \omega \int_{-\infty}^{\infty} P(x-y,t)\rho_a(y)\, dy \right) dx \\
&\quad -
\int_{-\infty}^{\infty} 
\phi\!\left( \omega \int_{-\infty}^{\infty} P(x,t)\rho_a(y)\, dy \right) dx ,
\end{aligned}
\end{equation}

\begin{equation}
    =  \omega\int_{-\infty}^{\infty} 
\!\left(  \int_{-\infty}^{\infty}\phi(x) P(x-y,t)\, dx \right)\rho_a(y) dy-~\omega
\int_{-\infty}^{\infty} 
\phi(x)~   P(x,t)~ dx .
\end{equation}
Substituting $x-y = z$ in the first integral leads to
\begin{equation}
    =  \omega\int_{-\infty}^{\infty} 
\!\left(  \int_{-\infty}^{\infty}\phi(y+z) P(z,t)\, dz \right)\rho_a(y) dy-
 \omega\int_{-\infty}^{\infty} 
\phi   P(x,t) dx ,
\end{equation}

\begin{equation}
    =  \omega\int_{-\infty}^{\infty} 
P(z,t)\!\left(  \int_{-\infty}^{\infty}\phi(y+z) \rho_a(y)\, dy \right) dz -
\omega\int_{-\infty}^{\infty} 
\phi   P(x,t) dx .
\end{equation}

In the first integral, $z$ is a dummy variable; replacing $z$ by $x$ gives
\begin{equation}
    =  \omega\int_{-\infty}^{\infty} 
P(x,t)\!\left(  \int_{-\infty}^{\infty}\phi(x+y) \rho_a(y)\, dy \right) dx -
\omega\int_{-\infty}^{\infty} 
P(x,t)  \phi(x)  ~dx .
    \label{PSN term in adjoint relation}
\end{equation}
From relations (\ref{adjoint relation}), (\ref{Drift term in adjoint relation}), (\ref{Diffusion term in adjoint relation}), and (\ref{PSN term in adjoint relation}), the Backward Fokker-Planck operator follows
\begin{equation}
  \mathcal{L}_{\mathrm{BFP}}\phi(x,t) = A(x) \frac{\partial \phi(x,t)}{\partial x} + D\frac{\partial^2 \phi(x,t)}{\partial x^2} + \omega \int_{-\infty}^{\infty} [\phi(x+y,t) - \phi(x,t)]\rho_a(y)\, dy.
\end{equation}
which is Eq. (\ref{eq:backward f-p operator}) written for the transition probability.

\section{Derivation of the MFPT}
\label{appendix:particle_derivation}
Here, we give details of the calculations used in calculating the MFPT discussed in Section \ref{subsec:particle-escpae-rate} while deriving Eqs. (\ref{eq:the MFPT_Integral_form_3}), (\ref{eq:the MFPT_Integral_form_small_limit_3}), and (\ref{eq:the MFPT_Integral_form_large_limit}). For $x_m = 1$, Eq. (\ref{eq:the MFPT_Integral_form}) becomes,
\begin{equation}
    \begin{aligned}
    \mathcal{T}(x_0 = -x_m) = \frac{1}{D_{eff}}\int_{-x_m}^{x_m} dx~ e^{(x^4-2x^2)/D_{eff}}\int_{-\infty}^{0}dx'~e^{-(x'^4-2x'^2)/D_{eff}} +\\ \frac{1}{D_{eff}}\int_{-x_m}^{x_m} dx~ \mathcal{I}(x)e^{(x^4-2x^2)/D_{eff}},
    \end{aligned}
    \label{eq:the MFPT_Integral_form_2_appendix}
\end{equation}
where $\mathcal{I}(x) = \int_{0}^{x}dx'e^{-(x'^4-2x'^2)/D_{eff}} $, since $\mathcal{I}(x)$ is an odd function of $x$, so the last integral does not contribute. Substituting $x' = -x'$, gives,
\begin{equation}
    \mathcal{T}(x_0 = -x_m) = \frac{1}{D_{eff}}\int_{-x_m}^{x_m} dx~ e^{(x^4-2x^2)/D_{eff}}\int_{0}^{\infty}dx'~e^{-(x'^4-2x'^2)/D_{eff}}.
    \label{eq:the MFPT_Integral_form_2_1_appendix}
\end{equation}
We use the integral $\int_0^\infty exp(-a^2x^4-2b^2x^2)~dx = 2^{-3/2}\frac{b}{a} e^{\frac{b^4}{2a^2}} K_{1/4} \left( \frac{b^4}{2a^2} \right)$ with analytic continuation $K_v(ze^{m\pi i}) = e^{-mv\pi i} K_v(z) - \pi i \frac{sin(mv\pi)}{sin(v\pi)}I_v(z)$, where $I_v(z)$ and $K_v(z)$ are modified Bessel functions of the first kind and second kind, respectively \cite{glasser2012integrals}. The expression for the MFPT becomes
\begin{equation}
    \begin{aligned}
    \mathcal{T}(x_0 = -x_m) = \frac{1}{D_{eff}} 2^{-3/2} e^{\frac{1}
    {2D_{eff}}}\left( K_{1/4} \left(\frac{1}{2D_{eff}}\right ) + \pi \sqrt{2}
    I_{1/4} \left(\frac{1}{2D_{eff}}\right )\right) \\ \int_{-x_m}^{x_m} dx~ e^{(x^4-2x^2)/D_{eff}}.
    \end{aligned}
    \label{eq:the MFPT_Integral_form_appendix}
\end{equation}  
In the weak noise regime $(D_{eff} <<1)$, arguments of modified Bessel functions become very large. Asymptotic expansions for large argument $z$ of modified Bessel functions are $I_v(z) = \frac{e^z}{\sqrt{2\pi z}}\left( 1- \frac{4v^2-1}{8z} + ...\right)$ and $K_v(z) = \sqrt{\frac{\pi}{2 z}} e^{-z}\left( 1+ \frac{4v^2-1}{8z} + ...\right)$ \cite{abramowitz1948handbook}. Keeping only the leading order term of the expansion gives
\begin{equation}
    \mathcal{T}(x_0 = -x_m) \approx \frac{1}{D_{eff}} 2^{-3/2}\sqrt{\pi D_{eff}} \left(  1 + \sqrt{2}e^{\frac{1}{D_{eff}}}  \right) \int_{-x_m}^{x_m} dx~ e^{(x^4-2x^2)/D_{eff}}.
    \label{eq:the MFPT_Integral_form_small_limit_appendix}
\end{equation}
In this limit, the integrand $e^{(x^4-2x^2)/D_{eff}}$ is sharply peaked at $x = 0$; therefore, the method of steepest descent can be used to solve this integral. The integral can be approximated by substituting $x = y\sqrt{D_{eff}}$, giving 
\begin{equation}
    \int_{-x_m}^{x_m} dx~ e^{(x^4-2x^2)/D_{eff}} = \sqrt{D_{eff}}\int_{-x_m/D_{eff}}^{x_m/D_{eff}} dy~ e^{-2y^2 + y^4D_{eff}} ,
\end{equation} 
in the $D_{eff} \rightarrow 0$ limit, it becomes \cite{gardiner2004handbook}, 
\begin{equation}
    \sqrt{D_{eff}}\int_{-x_m/D_{eff}}^{x_m/D_{eff}} dy~ e^{-2y^2 + y^4D_{eff}}  \approx \sqrt{D_{eff}} \int_{-\infty}^{\infty} dy~ e^{-2y^2} = \sqrt{\frac{\pi D_{eff}}{2}}.
\end{equation}
Keeping the leading order term, Eq. (\ref{eq:the MFPT_Integral_form_small_limit_appendix}) gives,
\begin{equation}
    \mathcal{T}(x_0 = -x_m) \approx \frac{\pi}{2\sqrt{2}} e^{1/D_{eff}}.
    \label{eq:the MFPT_Integral_form_small_limit_appendix_1}
\end{equation}

In the strong noise regime $(D_{eff} >>1)$, the arguments of modified Bessel functions become very small. In the limit $z \rightarrow 0$ modified Bessel function can be approximated as $I_v(z) \sim \left( \frac{1}{2}z \right)^v/\Gamma(v+1)$ and $K_v(z) \sim \frac{1}{2} \Gamma(v) \left( \frac{1}{2}z \right)^{-v} + \frac{1}{2} \Gamma(-v) \left( \frac{1}{2}z \right)^{v}$, here, $\Gamma$ represents the gamma function \cite{abramowitz1948handbook}. The integral $\int_{-x_m}^{x_m} dx e^{(x^4-2x^2)/D_{eff}}$ in the MFPT expression for large $D_{eff}$ can be approximated to $2x_m$. Thus, the expression can be written as
\begin{equation}
    \mathcal{T}(x_0 = -x_m) \approx  \frac{x_m}{2} \left( \Gamma\left(\frac{1}{4}\right) (D_{eff})^{-3/4} -\frac{1}{2}\Gamma\left(-\frac{1}{4}\right) (D_{eff})^{-5/4}\right).
    \label{eq:the MFPT_Integral_form_large_limit_appendix}
\end{equation}

\section{Other Distributions of Kick Amplitudes}
\label{nongauss}
In the main text, a Gaussian distribution for kick amplitudes is used. In small variance approximation, the Gaussian distributed PSN kicks can be equivalently represented by the active diffusion term $D_a$. Here, we investigate the MFPT for different kick amplitude distributions in different variance limits. The active diffusion term for any symmetric distribution can be obtained by Eq. (\ref{eq:the MFPT_detailed_II}). One obtains $D_a = \omega \sigma_a^2/2$ irrespective of the functional form of the symmetric distribution. This implies that PSN with any symmetric kick distribution can be represented as thermal noise under a small-variance approximation.
Other distributions, namely uniform, bimodal, and Laplace, their variances and respective active diffusion terms are listed in Table \ref{tab:distributions}.
\begin{figure}[!t]
    \centering
    \includegraphics[width=\textwidth]{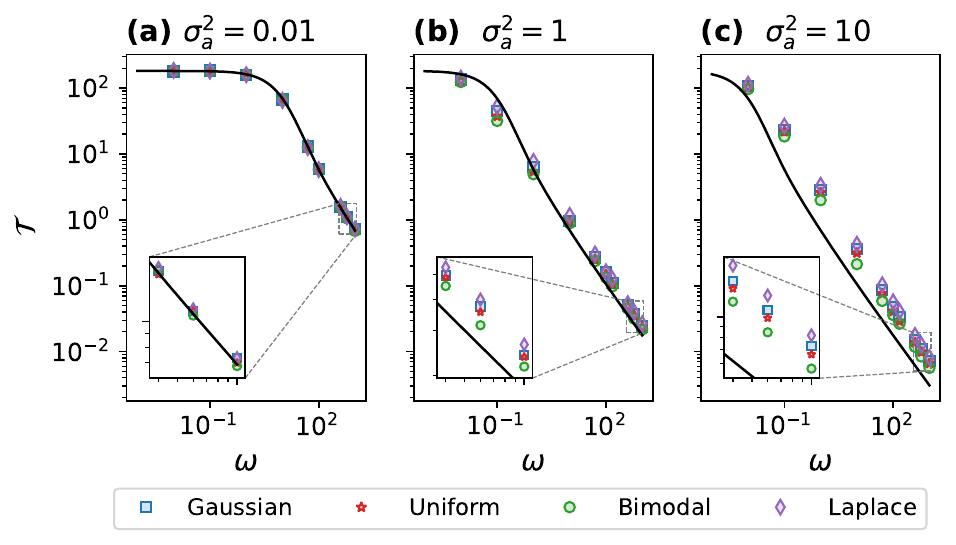}
    \caption{The MFPT $\mathcal{T}$ as a function of kick rate $\omega$ is plotted for different variance values \textbf{(a)} $\sigma_a^2 = 0.01$, \textbf{(b)} $\sigma_a^2 = 1$, and \textbf{(c)} $\sigma_a^2 = 10$. Solid line is the plot of Eq. (\ref{eq:the MFPT_Integral_form_3}) with $D_{eff} = D+D_a$. Simulation results match the theory in the small-variance case. Simulation parameters are $D = 0.2$, $\Delta t = 10^{-5}$ with an ensemble average of $2 \times10^{5}$. }
    \label{fig:Figure 13}
\end{figure}

\begin{table}[h]
\centering
\begin{tabular}{|c|c|c|}
\hline
Distribution ($\rho_a(y)$) & ~Variance ($\sigma_a^2$)~ & ~Active Diffusion ($D_a$) ~\\

\hline
$\frac{1}{2a}$ for $-a<y<a$, and $0$ otherwise&  $\frac{a^2}{3}$ & $\frac{\omega a^2}{6}$ \\
\hline
$\frac{1}{2}\left[\delta(y-a) + \delta(y+a)\right]$& $a^2$ & $\frac{\omega a^2}{2}$ \\
\hline
$\frac{1}{2a} e^{-\frac{|y|}{a}}$ & $2a^2$ & $\omega a^2$ \\
\hline
$\frac{1}{\sqrt{2\pi a^2}} e^{-\frac{y^2}{2a^2}}$ & $a^2$ & $\frac{\omega a^2}{2}$  \\
\hline
\end{tabular}
\caption{PSN kick-amplitude distributions, their variances, and corresponding calculated active diffusion coefficients under small variance approximation. The Gaussian distribution in the last row is given for reference.}
\label{tab:distributions}
\end{table}

In Fig. \ref{fig:Figure 13}, the MFPT versus kick rate for different kick-amplitude distributions is plotted. Fig. \ref{fig:Figure 13} (a) shows that all the distributions satisfy Eq. (\ref{eq:the MFPT_Integral_form_3}) satisfactorily, implying non-Gaussian noise-induced MFPT can be approximated as thermal noise-induced MFPT, whereas the deviation can be seen in sub-figures (b) and (c) for large variances. The MFPT is smallest for bimodal kicks, followed by uniform, Gaussian, and Laplace distributions.

\section{Numerical Simulations}
\label{simulate}
The Langevin equations can be solved using the standard Euler-Maruyama Method \cite{kloeden1992stochastic, perez2024brownian}. In this method, thermal noise was generated from a Wiener process, and to accommodate PSN into the algorithm, we approximate Poisson-shot noise as a Bernoulli process in the limit of a sufficiently small time step. Therefore, the position of the particle at time $t+\Delta t$ from the previous position at time $t$ can be calculated using the following algorithm,
\begin{equation}
    x(t+\Delta t) = x(t) + 4(x(t) - x^3(t)) \Delta t + \sqrt{2D} ~\mathcal{N}(0, \Delta t) + Y~ \mathrm{B}(\omega \Delta t),
\end{equation}
where $\mathcal{N}(0, \Delta t)$ is a Gaussian random variable with mean zero and variance $\Delta t$, and  $\mathrm{B}(\omega \Delta t)$ is the Bernoulli random variable that allows a kick with probability $\omega \Delta t$. $Y$ is the random variable with kick distribution $\rho_a(y)$ given by Eq. (\ref{eq:gaussian_psn_normalized_kick}). Eq. (\ref{eq:the MFPT_detailed}) can be solved numerically by writing the BFPE operator in matrix form and solving the matrix equation using the built-in Python functions. All simulations were performed using the NumPy library in Python \cite{harris2020array}, and the data were visualized with the Matplotlib library.


\bibliographystyle{apsrev}
\bibliography{ref}

\end{document}